\documentclass[reqno,11pt]{amsart}
\usepackage[utf8]{inputenc}
\usepackage{enumitem}
\usepackage[linktocpage]{hyperref}
\usepackage{graphicx}
\usepackage{amscd}
\usepackage{slashed}
\usepackage{amssymb}
\usepackage{mathtools} 
\usepackage[off]{auto-pst-pdf} 
\usepackage{tikz}
\usepackage{listings}
\usepackage{xcolor}
\usepackage[mathscr]{eucal}
\numberwithin{equation}{section}
\allowdisplaybreaks[1]

\title[Numerical Analysis of the Causal Action Principle]{Numerical Analysis of the Causal Action Principle
in Low Dimensions}

\author[F.\ Finster]{Felix Finster}
\address{Fakult\"at f\"ur Mathematik \\ Universit\"at Regensburg \\ D-93040 Regensburg \\ Germany}
\email{finster@ur.de}
\author[R.H. Jonsson]{Robert H. Jonsson}
\address{Max Planck Institute of Quantum Optics \\ Hans-Kopfermann-Str. 1 \\
D-85748 Garching \\ Germany}
\address{Nordita, KTH Royal Institute of Technology and Stockholm University, Hannes Alfv\'ens v\"ag 12, SE-106 91 Stockholm, Sweden}
\email{robert.jonsson@su.se}
\author[N. Kilbertus]{Niki Kilbertus \\ \\ January 2022}
\address{Department of Informatics \\ Technical University of Munich \\ Boltzmannstr. 3 \\ D-85748 Garching \\ Germany}
\address{Helmholtz AI \\ Ingolst\"adter Landstraße 1 \\ D-85764 Neuherberg \\Germany}
\email{niki.kilbertus@tum.de}

\newtheorem{Def}{Definition}[section]

\newtheorem{Prp}[Def]{Proposition}
\newtheorem{Lemma}[Def]{Lemma}

\newcommand{\Thanks}{\vspace*{.5em} \noindent \thanks}
\newcommand{\beq}{\begin{equation}}
\newcommand{\eeq}{\end{equation}}
\newcommand{\Proof}{\begin{proof}}
\newcommand{\QED}{\end{proof} \noindent}

\newcommand{\la}{\langle}
\newcommand{\ra}{\rangle}

\newcommand{\C}{\mathbb{C}}
\newcommand{\R}{\mathbb{R}}
\newcommand{\1}{\mbox{\rm 1 \hspace{-1.05 em} 1}}

\newcommand{\N}{\mathbb{N}}
\renewcommand{\H}{\mathscr{H}}
\renewcommand{\O}{\mathscr{O}}

\newcommand{\F}{{\mathscr{F}}}

\newcommand{\D}{{\mathscr{D}}}

\renewcommand{\O}{{\mathscr{O}}}
\renewcommand{\L}{{\mathcal{L}}}
\newcommand{\Sact}{{\mathcal{S}}}

\newcommand{\Lin}{\text{\rm{L}}}

\newcommand{\bitem}{\begin{itemize}[leftmargin=2.5em]}
\newcommand{\eitem}{\end{itemize}}

\DeclareFontFamily{OT1}{rsfso}{}
\DeclareFontShape{OT1}{rsfso}{m}{n}{ <-7> rsfso5 <7-10> rsfso7 <10-> rsfso10}{}
\DeclareMathAlphabet{\myscr}{OT1}{rsfso}{m}{n}

\newcommand\Felix[1]{}
\newcommand\Niki[1]{}
\newcommand\Robert[1]{}

\DeclareMathOperator{\tr}{tr}

\DeclareMathOperator{\diag}{diag}

\DeclareMathOperator{\supp}{supp}

\newcommand{\T}{\mathcal{T}}

\newcommand{\ii}{ i }

\newcommand{\braket}[2]{\left\langle \vphantom{#2} {#1}\left|\vphantom{#1}{#2}\right.\right\rangle}
\newcommand{\ketbra}[2]{\left| {#1}\vphantom{#2} \right\rangle\!\left\langle {#2}\vphantom{#1} \right|}

\definecolor{codegreen}{rgb}{0,0.6,0}
\definecolor{codegray}{rgb}{0.5,0.5,0.5}
\definecolor{codepurple}{rgb}{0.58,0,0.82}
\definecolor{backcolour}{rgb}{0.95,0.95,0.92}

\lstdefinestyle{mystyle}{
    backgroundcolor=\color{backcolour},   
    commentstyle=\color{codegreen},
    keywordstyle=\color{magenta},
    numberstyle=\tiny\color{codegray},
    stringstyle=\color{codepurple},
    basicstyle=\ttfamily\footnotesize,
    breakatwhitespace=false,         
    breaklines=true,                 
    captionpos=b,                    
    keepspaces=true,                 
    numbers=left,                    
    numbersep=5pt,                  
    showspaces=false,                
    showstringspaces=false,
    showtabs=false,                  
    tabsize=2
}

\begin{document}

\maketitle

\begin{abstract}
The numerical analysis of causal fermion systems is advanced by employing differentiable programming methods. The causal action principle for weighted counting measures is introduced for general values of the integer parameters~$f$ (the particle number), $n$ (the spin dimension) and $m$ (the number of spacetime points). In the case~$n=1$, the causal relations are clarified geometrically in terms of causal cones. Discrete Dirac spheres are introduced as candidates for minimizers for large~$m$ in the cases~$n=1, f=2$ and~$n=2, f=4$. We provide a thorough numerical analysis of the causal action principle for weighted counting measures for large~$m$ in the cases~$n=1,2$ and~$f=2,3,4$. Our numerical findings corroborate that all minimizers for large~$m$ are good approximations of the discrete Dirac spheres. In the example~$n=1, f=3$ it is explained how numerical minimizers can be visualized by projected spacetime plots. Methods and prospects are discussed to numerically investigate settings in which hitherto no analytic candidates for minimizers are known.
\end{abstract} 

\tableofcontents
\section{Introduction}
The theory of {\em{causal fermion systems}} is a recent approach to fundamental physics
(see the reviews~\cite{dice2014, review, dice2018}, the textbooks~\cite{cfs, intro} or the website~\cite{cfsweblink}).
In this approach, spacetime and all objects therein are described by a measure~$\rho$
on a set~$\F$ of linear operators of a Hilbert space~$(\H, \la .|. \ra_\H)$.
The physical equations are formulated via the so-called {\em{causal action principle}},
a nonlinear variational principle where an action~$\Sact$ is minimized under variations of the measure~$\rho$.

While the causal action principle has been analyzed in detail from an abstract point of view,
only very few examples of exact minimizers are known. In~\cite{small, support, bierler} a few simple minimizers were
constructed numerically in low dimensions and for a small number of spacetime points.
Moreover, the Dirac sea vacuum used as the starting point for most applications
is known to be a minimizer, but only in a specific limiting case when the ultraviolet regularization is removed
(for details see~\cite[Chapters~3-5]{cfs}).
In view of the lack of explicit examples, it is an important task to minimize the causal action
numerically for many spacetime points and, eventually, in a high-dimensional Hilbert space,
with the goal of clarifying the structure of the minimizing measures.
The present paper is a first step towards this goal, where we consider Hilbert spaces
of low dimensions two, three and four.

In the causal action principle, one minimizes the causal action under variations
of a measure~$\rho$ within the class of all regular Borel measures, under three constraints:
the volume, the trace and the boundedness constraints.
In order to describe the measure numerically, we here restrict our attention to {\em{weighted counting
measures}}~$\rho$ of the form
\[ \rho = \sum_{i=1}^m c_i \,\delta_{x_i} \:, \]
where~$\delta_{x_i}$ is the Dirac measure supported at~$x_i \in \F$,
and the coefficients~$c_i$ are non-negative weight factors.
Choosing~$m$ large enough, one can approximate any regular Borel measure by such
weighted counting measures. Clearly, in the numerical study one needs to choose~$m$ finite.
But information on general minimizing measures can be obtained by considering the asymptotics for
large~$m$. The volume and trace constraints can be implemented by considering
{\em{normalized}} measures on operators of {\em{fixed trace}}, i.e.
\[ \sum_{i=1}^m c_i = 1 \qquad \text{and} \qquad 
\text{$\tr x_i = 1$ for all~$i=1,\ldots, m$}\:. \]
The boundedness constraint, on the other hand, is needed in order to ensure that the
points~$x_i$ remain bounded and the infimum of the
causal action remains strictly positive in the limit~$m \rightarrow \infty$.
Since~$m$ will be fixed when minimizing numerically, here we may leave out the boundedness constraint.
Instead, the necessity of the boundedness constraint will become apparent in the fact that the
causal action found numerically in the case~$n=2$ will tend to zero asymptotically for large~$m$.

Minimizing the causal action principle can be regarded as a problem of nonlinear optimization.
The number of parameters grows linearly in the dimension~$f$ of the Hilbert space.
Therefore, the numerical analysis becomes more difficult the larger the dimension~$f$ of the Hilbert space is
chosen. For this reason, we here focus on the low-dimensional cases
\[ f=2, 3 \text{ and } 4 \:, \]
with the main goal of analyzing the asymptotics for large~$m$.
In the cases~$n=1, f=2$ and~$n=2, f=4$ it is conjectured that, asymptotically for large~$m$, the minimizing measures
should have the form of equally spaced points on a sphere of dimension two (if $f=2$) or four (if $f=4$),
where the radius of the sphere tends to infinity for large~$m$.
These so-called {\em{discrete Dirac sphere configurations}} will be introduced in detail in Sections~\ref{secdirac2d} and~\ref{secdirac4d}.
Our numerical results support the conjecture that such discrete Dirac sphere configurations are
indeed the only minimizing measures for large~$m$.

Our numerical findings also triggered a more detailed analytic study,
which gave some new insight into the structure of the causal action principle.
These analytic results will also be derived and explained in the present paper.
In particular, we realized that, in the case of spin dimension~$n=1$, the causal relations on~$\F$
give rise to cones. This is worked out in Sections~\ref{seclagcausal}
and~\ref{seccausalgen} (see also Figure~\ref{figlightcone} on page~\pageref{figlightcone}
and Figure~\ref{figlightcone2} on page~\pageref{figlightcone2}).
Moreover, the asymptotics for fixed~$m$ and large~$f$ was worked out and compared to the
numerical findings (see Section~\ref{secasyflarge}).

Our numerical investigations also include the case~$n=1$ and~$f=3$. In this case, it is unknown how the minimizing measures should look like for large~$m$.
Therefore, our numerical approach can give new insight into the structure of minimizing measures in this setting and inspire progress regarding analytical results.
To this end, in Section~\ref{secf3}, we introduce \emph{projected spacetime plots} as a tool to 
visualize the geometry and causal structure of a discrete spacetime, as seen from a reference spacetime point.

The paper is structured as follows. Section~\ref{secweighted} provides the necessary
background on causal fermion systems and the causal action principle.
Section~\ref{secintro2d} is devoted to the case~$f=2$.
After working out the Lagrangian and the causal structure in detail (Section~\ref{seclagcausal}),
we explain the underlying mechanism qualitatively (Section~\ref{secqualitative}).
We proceed by computing the causal action for spherically symmetric measures (so-called
isotropic measures; see Section~\ref{secisotropic}).
Then we show that the causal action can be decreased further by choosing discrete points
on a sphere (the so-called discrete Dirac sphere; see Section~\ref{secdirac2d}).
In these examples, the causal structure is trivial in the sense that all distinct
spacetime points are spacelike separated. Such measures are analyzed in some more detail
(Section~\ref{sectivcausal}).
In Section~\ref{secdirac4d} we turn our attention to the case~$n=2, f=4$ and 
generalize the discrete Dirac sphere to this case.
In Section~\ref{seccausalgen} the considerations on the causal structure from Section~\ref{seclagcausal}
are generalized to a general dimension~$f$ of the Hilbert space.
In Section~\ref{secnumerics} it is explained how we set up the problem numerically.
Section~\ref{secasyflarge} is devoted to the asymptotics for large~$f$ keeping~$m$ fixed.
In Section~\ref{secdiracnum} the cases~$n=1, f=2$ and~$n=2, f=4$
are investigated numerically. The numerical minimizers are compared to the discrete
Dirac sphere configurations.
In Section~\ref{secf3} we take the example~$n=1, f=3$ to discuss 
prospects for numerical studies in the case~$f > 2n$.
In Section~\ref{secconclusion} we conclude the paper with a brief summary and outlook.
Appendix~\ref{app:impdetails} provides some details of our numerical implementation.

\section{The Causal Action Principle for Weighted Counting Measures} \label{secweighted}
For clarity, we first introduce the causal action principle for weighted counting measures
and explain afterward the connection to the general causal action principle as introduced in~\cite[\S1.1.1]{cfs}.
For a more detailed introduction to the physical concepts and the mathematical structures
we refer to the textbooks~\cite{cfs, intro}.

Given a complex Hilbert space of finite dimension~$f$ and a parameter~$n \in \N$
(the ``spin dimension''), we let~$\F \subset \Lin(\H)$ be the set of all
symmetric\footnote{Here, by a symmetric operator~$A$ we mean that~$\la A u | v \ra_\H =
\la u | A v \ra_\H$ for all~$u,v \in \H$. Representing the operator in
an orthonormal basis, the resulting matrix is Hermitian.
For bounded operators as considered here,
the notions ``symmetric'' and ``selfadjoint'' coincide. }
 linear operators~$x$ on~$\H$ with trace one,
\beq \label{fixedtrace}
\tr x = 1 \:,
\eeq
which (counting multiplicities) have at most~$n$ positive and at most~$n$ negative eigenvalues.
Next, we let~$\rho$ be a {\em{normalized weighted counting measure}} on~$\F$
(as first considered in the context of causal fermion systems in~\cite[Section~2]{support}). 
Thus, given~$m \in \N$, we choose points~$x_1, \ldots, x_m \in \F$
and corresponding weights~$c_1, \ldots, c_m$ with
\beq \label{cnorm}
c_i \geq 0 \qquad \text{and} \qquad \sum_{i=1}^m c_i = 1\:.
\eeq
We introduce the measure~$\rho$ by defining its integral of a continuous function~$f$ by
\beq \label{weightcount}
\int_{\F} f\, d\rho := \sum_{i=1}^m c_i \,f(x_i) \qquad \text{for all~$f \in C^0(\F)$}\:.
\eeq
We also use the shorter notation
\beq \label{weightdelta}
\rho = \sum_{i=1}^m c_i \,\delta_{x_i} \:,
\eeq
where~$\delta_{x_i}$ is the Dirac measure supported at~$x_i$.

For any~$x, y \in \F$, the product~$x y$ is an operator
of rank at most~$2n$. We denote its non-trivial eigenvalues counting algebraic multiplicities
by~$\lambda^{xy}_1, \ldots, \lambda^{xy}_{2n} \in \C$
(more specifically,
denoting the rank of~$xy$ by~$k \leq 2n$, we choose~$\lambda^{xy}_1, \ldots, \lambda^{xy}_{k}$ as all
the non-zero eigenvalues and set~$\lambda^{xy}_{k+1}, \ldots, \lambda^{xy}_{2n}=0$).
We introduce the causal Lagrangian and the causal action by
\begin{align}
\text{\em{causal Lagrangian:}} && \L(x,y) &= \frac{1}{4n} \sum_{i,j=1}^{2n} \Big( \big|\lambda^{xy}_i \big| - \big|\lambda^{xy}_j \big| \Big)^2 \label{Lagrange} \\
\text{\em{causal action:}} && \Sact(\rho) &= \iint_{\F \times \F} \L(x,y)\: d\rho(x)\, d\rho(y) \:. \label{Sdef}
\end{align}
The {\em{causal action principle}} is to minimize~$\Sact$
by varying $\rho$ in the class of normalized weighted counting measures~\eqref{weightcount}.
The parameters entering this variational principle are
\[ \left\{ \;\;\begin{array}{cl}
n &\qquad \text{spin dimension} \\
m &\qquad \text{maximal number of points of weighted counting measure} \\
f &\qquad \text{dimension of the Hilbert space} \:.
\end{array} \right. \]

The spectral properties of the operator product~$xy$ gives rise to the following {\em{causal structure}}.
Two points~$x, y \in \F$ are called {\em{spacelike}} separated if all the eigenvalues~$\lambda^{xy}_j$
have the same absolute value.
They are said to be {\em{timelike}} separated if the~$\lambda^{xy}_j$ are all real and do not all 
have the same absolute value.
In all other cases (i.e.\ if the~$\lambda^{xy}_j$ are not all real and do not all 
have the same absolute value), the points~$x$ and~$y$ are said to be {\em{lightlike}} separated.
Evaluating these relations for the points~$x_1, \ldots, x_m$, we get corresponding causal relations
between these points, which thus form our {\em{spacetime}}~$M$,
\beq \label{Mdef}
M := \{ x_1, \ldots, x_m\} \subset \F \:.
\eeq
The causal action principle is compatible with the above causal structure in the
following sense. Suppose that two spacetime points~$x,y \in M$ are spacelike separated.
Then the eigenvalues~$\lambda^{xy}_i$ all have the same absolute value,
implying that the Lagrangian~\eqref{Lagrange} vanishes.
Working out the corresponding Euler-Lagrange equations (for details 
see~\cite{jet} or~\cite[Chapter~7]{intro}), one finds that pairs of points with spacelike separation again drop out.
This can be seen in analogy to the usual notion of causality where
points with spacelike separation cannot influence each other.
In this sense, the principle of causality is built into the theory of causal fermion systems.

We conclude by explaining how the above variational principle is obtained from the general causal action principle
introduced in~\cite[\S1.1.1]{cfs}. Before beginning, we point out that, in order to allow for a numerical treatment,
we clearly need to assume that~$\H$ is finite-dimensional and that~$\rho$ is a weighted counting measure.
Moreover, using the rescaling freedom $\rho \rightarrow \sigma \rho$, it is no loss of generality to restrict
attention to normalized measures.
Next, using that minimizing measures are supported on operators of constant trace
(see~\cite[Proposition~1.4.1]{cfs}), we may fix the trace of the operators. Moreover, 
by rescaling the operators according to~$x \rightarrow \lambda x$ with~$\lambda \in \R$,
one can assume without loss of generality that this trace is equal to one~\eqref{fixedtrace}.
Finally, as already mentioned in the introduction,
we here disregard the so-called {\em{boundedness constraint}}, which
states that all the measures~$\rho$ taken into account in the minimization process
should satisfy the inequality
\beq \label{Tdef}
\T(\rho) := \iint_{\F \times \F} \bigg( \sum_{i=1}^{2n} \big|\lambda^{xy}_i \big| \bigg)^2\: d\rho(x)\, d\rho(y) \leq C
\eeq
for a given parameter~$C>0$.
The boundedness constraint is needed for the causal action principle to be well-posed
when varying within the class of all regular Borel measures. Specifically, in that case, it ensures compactness.
However, in our setting of weighted counting measures
with an upper bound~$m$ on the number of spacetime points,
the boundedness constraint is {\em{not needed}} for the causal action principle
to be well-posed. Nevertheless, the functional~$\T$ will be relevant for us when analyzing
the asymptotics as~$m \rightarrow \infty$. Indeed, in this limiting case, the
support of the minimizing measures will typically tend to infinity.
Likewise, the functional~$\T$ will tend to infinity.
This phenomenon will be discussed in more detail at the end of Section~\ref{secdirac2d} and in Section~\ref{secdirac4d}.

\section{The Causal Action Principle on a Two-Dimensional Hilbert Space} \label{secintro2d}
The simplest interesting setting of the causal action principle is to choose~$n=1$ and~$f=2$.
In this section we analyze this case in detail. This will lead us to the conjecture that, for large values of~$m$,
minimizing measures should be close to so-called discrete Dirac sphere configurations
(as will be introduced in Section~\ref{secdirac2d}). This section also sets the stage for the
numerical analysis in Section~\ref{secnumerics}, in which this conjecture will be examined.

We identify~$\H$ with~$(\C^2, \la .,. \ra_{\C^2})$ (where~$\la .,. \ra_{\C^2}$ denotes the
standard Euclidean scalar product).
Then the symmetric operators on~$\H$ are identified with the Hermitian $2 \times 2$-matrices.
Every such matrix can be represented as a real linear combination of the identity matrix and the
three Pauli matrices 
\beq \label{pauli}
\sigma^1 = \begin{pmatrix} 0 & 1 \\ 1 & 0 \end{pmatrix} , \qquad
\sigma^2 = \begin{pmatrix} 0 & -i \\ i & 0 \end{pmatrix} \qquad \text{and} \qquad
\sigma^3 = \begin{pmatrix} 1 & 0 \\ 0 & -1 \end{pmatrix}\:.
\eeq
Prescribing the trace~\eqref{fixedtrace} and imposing that the matrix should have at most one
positive and at most one negative eigenvalue, one readily finds that the operators in~$\F$
can be identified with the complement of the open ball in~$\R^3$ via the mapping
\beq\label{eq:R3DiracMap}
F \::\: [1,\infty)\times S^2 \to \F,\quad (\tau,\vec x)\mapsto F_\tau(\vec x) = \frac12\left(\1+\tau \sum_{i=1}^3 x^i \sigma^i\right) := \frac{1}{2}\: \big( \1+\tau \vec x\cdot\vec \sigma \big)
\eeq
(here~$S^2$ denotes the unit sphere in~$\R^3$).
Indeed, the matrix~$F_\tau(\vec x)$ clearly is Hermitian and has trace one.
Conversely, every Hermitian matrix of trace one is the unique image of some~$(\tau,\vec x)\in(1,\infty)\times S^2$.
Next, the matrix~$F_\tau(\vec x)$ satisfies the polynomial equation
\[ \big( 2 F_\tau(\vec x) - \1 \big)^2 = \bigg( \sum_{i=1}^3 \tau\: x^i \sigma^i \bigg)^2 = \tau^2\:\1 \:, \]
where in the last step we used the anti-commutation relations for Pauli
matrices $\{\sigma^i, \sigma^j\} = 2 \delta^{ij}$. Consequently, the eigenvalues of~$F_\tau(\vec x)$ are given by
\beq \label{nupmform}
\nu_\pm = \frac{1}{2} \:\big( 1 \pm \tau \big) \:.
\eeq
In particular, as desired, $F_\tau(\vec x)$ has at most one positive and at most one negative eigenvalue.

To simplify notation, in what follows we use the map~$F$ in~\eqref{eq:R3DiracMap}
to identify~$\F$ with the set~$[1, \infty) \times S^2$. We also write~$F_\tau(\vec x)$
simply as~$(\tau, \vec x) \in \F$. 

\subsection{The Lagrangian and the Causal Structure} \label{seclagcausal}
Before analyzing the causal action principle,
we need to compute the Lagrangian for two general arguments in~$\F$.

\begin{Lemma} \label{lemmalagrangianR3}
Let $\tau,\tau'\in[1,\infty)$ and~$\vec x,\vec{x}'\in S^2$. We denote the angle between~$\vec{x}$
and~$\vec{x}'$ by~$\vartheta \in [0, \pi]$ (i.e.\ $\cos \vartheta = \vec{x}\cdot \vec{y}$).
Then the operator product $F_\tau(\vec x)F_{\tau'}(\vec x')$ has the eigenvalues
\beq\label{evals_product}
\lambda_{1,2}=\frac{1}{4}\: \Big(1+\tau\tau'\cos\vartheta \pm\sqrt{  (1+\tau \tau'\cos \vartheta)^2-(\tau^2-1)(\tau'^2-1)  }\Big) \:.
\eeq
Moreover,
\beq\label{lagrangian_R3}
\L \big((\tau, \vec x), (\tau', \vec x') \big) = \frac1{8} \Big( \big( 1+\tau\tau'\cos\vartheta \big)^2-(\tau^2-1)(\tau'^2-1) \Big) \,\big( 1- \chi_{[\vartheta_-,\vartheta_+]}(\vartheta) \big) \:,
\eeq
where $ \chi_{[\vartheta_-,\vartheta_+]}$ is the characteristic function of the interval defined by
\beq\label{eq:critical_angles}
\cos\vartheta_\pm=\frac{-1\mp\sqrt{(\tau^2-1)(\tau'^2-1)}}{\tau\tau'} \:.
\eeq
\end{Lemma}
\Proof 
The proof is a direct generalization of the computations in the case $\tau=\tau'$ as carried out
in~\cite[Proposition~4.1]{small} and~\cite[Example~2.8]{continuum}.
Using the identity between Pauli matrices
\[ 
\sigma^j \sigma^k = \delta^{jk}\:\1 + i\: \epsilon^{jkl}\: \sigma^l \]
(where~$\epsilon^{jkl}$ is the totally anti-symmetric Levi-Civita symbol),
one obtains
\beq \label{FF}
F_\tau(\vec x) \,F_{\tau'}(\vec x')
= \frac14 \: \Big( (\tau\tau'\cos\vartheta+1) \1 + \big(\tau\vec x+\tau'\vec x'+\ii\tau\tau' \vec x\wedge\vec x' \big)\cdot\vec \sigma \Big)
\eeq
(where the wedge product is defined by $(\vec{x} \wedge \vec{y})^l =
\epsilon^{jkl} x^j y^k$).
Using that the vector~$\vec{x} \wedge \vec{x}'$ is orthogonal to
both~$\vec{x}$ and~$\vec{x}'$, a direct computation using the transformation
\[ \tau^2+\tau'^2 + 2 \tau \tau' \cos \vartheta - \tau^2 \tau'^2 \sin^2 \vartheta = (1+\tau \tau'\cos \vartheta)^2-(\tau^2-1)(\tau'^2-1) \]
shows that
\[ \Big( F_\tau(\vec x)F_{\tau'}(\vec x') -\frac14 \:\big(\tau\tau'\cos\vartheta+1 \big)\,\1\Big)^2
= \frac1{16}\: \Big( \big(1+ \tau \tau'\cos\vartheta \big)^2-(\tau^2-1)(\tau'^2-1) \Big)\, \1 \:. \]
The eigenvalues of~$F_\tau(\vec x)F_{\tau'}(\vec x')$ are the roots of this polynomial.
This gives~\eqref{evals_product}.

If the vectors~$\vec{x}$ and~$\vec{x}'$ are collinear, these eigenvalues simplify to
\[ \lambda_{1,2}=  \left\{ \begin{array}{ll} \displaystyle \frac14\,(1\pm \tau)(1\pm \tau')>0 & \text{if~$\vartheta=0$} \\[0.9em]
\displaystyle \frac14\, (1\pm \tau)(1\mp \tau')<0 & \text{if~$\vartheta=\pi $} \:.
\end{array} \right. \]
In both cases, the eigenvalues are real. On the other hand, if the argument of the square root in~\eqref{evals_product}
is negative, then the eigenvalues form a complex conjugate pair.
This is the case if 
\[ \frac{-1-\sqrt{(\tau^2-1)(\tau'^2-1)}}{\tau \tau'}<\cos\vartheta<\frac{-1+\sqrt{(\tau^2-1)(\tau'^2-1)}}{\tau \tau'} \:. \]
Hence $\L(F_\tau(\vec x),F_{\tau'}(\vec x'))>0$ only if the angle $\vartheta$ lies outside this interval. In this case,
the relation
\beq \label{prodsign}
\lambda_1 \lambda_2=\frac1{16}(\tau^2-1)(\tau'^2-1)>0
\eeq
shows that the eigenvalues have the same sign. Therefore,
\beq \label{Lform2}
\L \big( (\tau, \vec x), (\tau', \vec x') \big) = \frac12 \,\big(|\lambda_1|-|\lambda_2| \big)^2
= \frac{1}{2} \,\big(\lambda_1 - \lambda_2 \big)^2 \:.
\eeq
Substituting the formula for the eigenvalues~\eqref{evals_product} gives~\eqref{lagrangian_R3}.
\QED

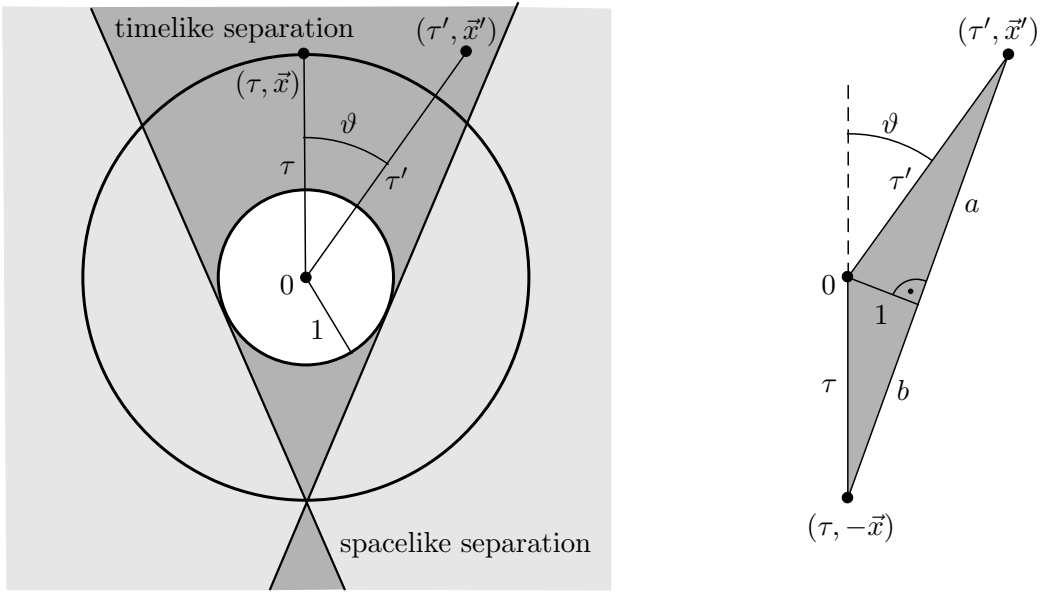
\begin{figure}
\psscalebox{1.0 1.0} 
{
\begin{pspicture}(0,26.14992)(13.455036,34.001755)
\definecolor{colour1}{rgb}{0.7019608,0.7019608,0.7019608}
\definecolor{colour0}{rgb}{0.9019608,0.9019608,0.9019608}
\pspolygon[linecolor=black, linewidth=0.02, fillstyle=solid,fillcolor=colour1](13.380036,33.275917)(11.250036,27.385918)(11.250036,30.325918)
\pspolygon[linecolor=colour0, linewidth=0.01, fillstyle=solid,fillcolor=colour0](1.1750367,33.915916)(4.045037,27.355919)(3.5600367,26.170918)(0.07003662,26.170918)(0.015036621,33.92592)
\pspolygon[linecolor=colour0, linewidth=0.01, fillstyle=solid,fillcolor=colour0](6.8500366,33.915916)(8.100037,33.89592)(8.100037,26.165918)(4.5900364,26.175919)(4.0800366,27.355919)
\pspolygon[linecolor=colour1, linewidth=0.01, fillstyle=solid,fillcolor=colour1](3.5850365,26.170918)(4.0550365,27.300919)(4.565037,26.180918)
\pspolygon[linecolor=colour1, linewidth=0.01, fillstyle=solid,fillcolor=colour1](3.0050366,29.720919)(4.065037,27.370918)(5.025037,29.680918)(4.8750367,29.490917)(4.7650366,29.370918)(4.4750366,29.210918)(4.2050366,29.140919)(3.8550367,29.150917)(3.5450366,29.250917)(3.2850366,29.420918)(3.1350367,29.570917)
\pspolygon[linecolor=colour1, linewidth=0.01, fillstyle=solid,fillcolor=colour1](1.1850367,33.915916)(6.8350368,33.92592)(5.2050366,30.085918)(5.2150364,30.535917)(5.0950365,30.825918)(4.8950367,31.135918)(4.6350365,31.345919)(4.2850366,31.485918)(3.9350367,31.505919)(3.5450366,31.405918)(3.2850366,31.215918)(3.0650365,30.965918)(2.9450366,30.745918)(2.8650367,30.445917)(2.8850367,30.035917)
\pscircle[linecolor=black, linewidth=0.04, dimen=outer](4.045037,30.325918){1.2}
\pscircle[linecolor=black, linewidth=0.04, dimen=outer](4.045037,30.325918){3.0}
\psline[linecolor=black, linewidth=0.03](4.565037,26.165918)(1.18837,33.90592)
\psline[linecolor=black, linewidth=0.03](3.5650365,26.165918)(6.86837,33.985916)
\pscircle[linecolor=black, linewidth=0.01, fillstyle=solid,fillcolor=black, dimen=outer](4.045037,30.325918){0.07}
\pscircle[linecolor=black, linewidth=0.01, fillstyle=solid,fillcolor=black, dimen=outer](4.0150366,33.30592){0.07}
\pscircle[linecolor=black, linewidth=0.01, fillstyle=solid,fillcolor=black, dimen=outer](6.1750364,33.33592){0.07}
\psline[linecolor=black, linewidth=0.02](4.0400367,30.335918)(4.6600366,29.305918)
\psline[linecolor=black, linewidth=0.02](4.0400367,30.315918)(6.190037,33.33592)
\psline[linecolor=black, linewidth=0.02](4.0400367,30.345919)(4.0200367,33.295918)
\psarc[linecolor=black, linewidth=0.02, dimen=outer](4.0400367,30.305918){1.88}{54.0}{91.0}
\pscircle[linecolor=black, linewidth=0.01, fillstyle=solid,fillcolor=black, dimen=outer](11.245037,30.335918){0.07}
\pscircle[linecolor=black, linewidth=0.01, fillstyle=solid,fillcolor=black, dimen=outer](13.385036,33.295918){0.07}
\pscircle[linecolor=black, linewidth=0.01, fillstyle=solid,fillcolor=black, dimen=outer](11.245037,27.395918){0.07}
\psline[linecolor=black, linewidth=0.02, linestyle=dashed, dash=0.17638889cm 0.10583334cm](11.255036,30.335918)(11.255036,32.885918)
\psarc[linecolor=black, linewidth=0.02, dimen=outer](11.270037,30.355919){1.88}{54.0}{91.0}
\psline[linecolor=black, linewidth=0.02](11.265037,30.320917)(12.195037,29.960918)
\psarc[linecolor=black, linewidth=0.02, dimen=outer](12.160037,29.965918){0.335}{67.0}{157.0}
\pscircle[linecolor=black, linewidth=0.01, fillstyle=solid,fillcolor=black, dimen=outer](12.085036,30.140919){0.03}
\rput[bl](3.7,30.1){$0$}
\rput[bl](4.1,29.5){$1$}
\rput[bl](3.7,31.7){$\tau$}
\rput[bl](4.5,32.25){$\vartheta$}
\rput[bl](5.1,31.5){$\tau'$}
\rput[bl](4.5,26.6){spacelike separation}
\rput[bl](1.5,33.45){timelike separation}
\rput[bl](10.9,30.1){$0$}
\rput[bl](11.7,32.25){$\vartheta$}
\rput[bl](3.1,32.7){$(\tau, \vec{x})$}
\rput[bl](5.5,33.4){$(\tau', \vec{x}')$}
\rput[bl](12.7,33.4){$(\tau', \vec{x}')$}
\rput[bl](10.7,26.8){$(\tau, -\vec{x})$}
\rput[bl](10.9,28.8){$\tau$}
\rput[bl](11.8,31.5){$\tau'$}
\rput[bl](12.8,31.2){$a$}
\rput[bl](11.9,28.7){$b$}
\rput[bl](11.6,29.7){$1$}
\end{pspicture}
}
\caption{Causal structure on~$\F$ in the case~$f=2$.}
 \label{figlightcone}
\end{figure}

We next analyze the {\em{causal relations}} as introduced abstractly before~\eqref{Mdef}.
According to the explicit formula~\eqref{evals_product}, the eigenvalues are either
both real and have the same sign (see~\eqref{prodsign}), or else they form a complex
conjugate pair. These two cases correspond precisely to {\em{timelike}} and {\em{spacelike}}
separation. In the boundary case between timelike and spacelike
we have a single degenerate eigenvalue. This occurs when the discriminant in~\eqref{evals_product}
vanishes, i.e.\ if
\beq \label{lightcone}
(1+\tau \tau'\cos \vartheta)^2 = (\tau^2-1)(\tau'^2-1) \:.
\eeq
This equation has a simple geometric interpretation shown in Figure~\ref{figlightcone}.
In this figure, the points parametrized by~$(\tau, \vec{x})$ and~$(\tau', \vec{x}')$ are shown,
and~$\vartheta$ is again the angle between~$\vec{x}$ and~$\vec{x}'$.
Given~$(\tau, \vec{x})$, all the points~$(\tau', \vec{x}')$ with timelike separation
form a region of the form of a three-dimensional double cone, with the ball of radius one removed.
The antipodal point~$(\tau, -\vec{x})$ is the apex of this cone, and its boundary is
tangential to the sphere of radius one. This causal structure can be derived using elementary trigonometry.
We here explain the main step, which consists in verifying that the boundary between timelike
and spacelike separation is formed by the boundary of the cone. To this end, we consider the
triangle formed of the origin,
the point~$(\tau', \vec{x}')$ and the antipodal point of~$(\tau, \vec{x})$ (see the right of Figure~\ref{figlightcone}).
We only consider the case~$\vartheta<\pi/2$, noting that the case~$\vartheta>\pi/2$ is analogous.
Applying the law of cosines to this triangle gives
\beq \label{eq1}
(a+b)^2 = \tau^2 + \tau'^2 - 2 \tau \tau' \cos(\pi - \vartheta) = \tau^2 + \tau'^2 + 2 \tau \tau' \cos \vartheta \:.
\eeq
On the other hand, the Pythagorean theorem yields~$a=\sqrt{\tau'^2-1}$ and~$b=\sqrt{\tau^2-1}$, and thus
\beq \label{eq2}
(a+b)^2 = \tau^2+\tau'^2 - 2 + 2 \sqrt{(\tau^2-1)(\tau'^2-1)} \:.
\eeq
Combining~\eqref{eq1} and~\eqref{eq2} gives~\eqref{lightcone}.

\subsection{Qualitative Description of Underlying Mechanisms} \label{secqualitative}
The causal structure shown in Figure~\ref{figlightcone} also gives a good intuitive
understanding of the causal action principle, as we now explain.
The Lagrangian in the causal action~\eqref{Sdef} 
can be regarded as a {\em{non-local potential}} describing a ``pair interaction'' of the points~$x,y \in M$
(with~$M$ according to~\eqref{Mdef}). Since the Lagrangian is non-negative, this
non-local potential is {\em{repulsive}}. Moreover, this non-local potential vanishes
unless the points are timelike separated.
Therefore, in simple terms, the causal action principle aims for reaching a configuration
where as many pairs of points as possible are spacelike separated.
Since the opening angle of the light cone becomes smaller when~$\tau$ is increased
(as one sees in Figure~\ref{figlightcone} if the outer circle is made large),
aiming for spacelike separation also leads to the tendency to increase~$\tau$ and~$\tau'$.
However, there is another competing effect: Every point~$x \in M$ also contributes to the causal action
via the pair interaction with itself. This {\em{self-contribution}} can be computed by evaluating~\eqref{lagrangian_R3}
in the case~$\tau'=\tau$ and~$\vec{x}'=\vec{x}$,
\beq \label{Ldiag}
\L \big((\tau, \vec x), (\tau, \vec x) \big) =
\frac1{8} \Big( \big( 1+\tau^2)^2-\big(\tau^2-1 \big)^2 \Big) = \frac{\tau^2}{2} \:.
\eeq
The resulting contribution to the causal action gets larger if~$\tau$ is increased.
As a consequence of these competing effects, the parameter~$\tau$ should get large, but not too
large.

This qualitative consideration conveys the correct intuition, but it does not explain
how minimizers should look like. To this end, we need finer and more quantitative arguments
as given in the next sections.

\subsection{Minimizing in the Class of Isotropic Measures} \label{secisotropic}
As a first step towards the quantitative analysis,
we want to shed light on the angular distribution of the points~$x_i$ for minimizing
configurations.
Since the variational principle is spherically symmetric, it is natural to expect that,
in the limit~$m \rightarrow \infty$, minimizing measures should also have this property.
We refer to measures with spherical symmetry as {\em{isotropic measures}}.
They can be written as
\beq \label{spherical}
d\rho = \frac{1}{4 \pi}\: d\mu(\tau) \:d\cos \vartheta\: d\varphi \:,
\eeq
where~$\mu$ is a normalized measure on~$[1, \infty)$,
\[ 
\int_1^\infty d\mu = 1 \]
(more precisely, $\mu$ should be a regular Borel measure; for technical simplicity,
the reader may restrict attention to the case~$d\mu = h(\tau)\, d\tau$ with~$h$
a non-negative continuous function).
Clearly, such a measure is {\em{not}} a weighted counting measure (see~\eqref{weightcount}),
but it can be approximated by such measures in the weak sense.
\begin{Prp}\label{propositionn1_fixedtau} For an isotropic measure~\eqref{spherical}, the causal action
can be written as
\begin{align}
\Sact(\rho) &= \int_1^\infty d\mu(\tau) \int_1^\infty d\mu(\tau') \: \L_\text{iso}(\tau, \tau') 
\qquad \text{with} \\
 \L_\text{\rm{iso}}(\tau, \tau') &= \frac{(\tau^2-1)^\frac{3}{2}\: (\tau'^2-1)^\frac{3}{2}}{12\, \tau \tau'} - \frac{\tau^2 \tau'^2}{12} + \frac{\tau^2+\tau'^2}{8} \:. \label{Liso}
\end{align}
Moreover, the isotropic Lagrangian~$\L_\text{\rm{iso}}$ is bounded from below by
\beq\label{eq:iso_measure_bound}
\L_\text{\rm{iso}}(\tau, \tau') > \L_\text{\rm{iso}}(1,1) = \frac{1}{6}\qquad \text{for all~$\tau, \tau'
\in [1, \infty)$ with~$(\tau, \tau') \neq (1,1)$}\:.
\eeq
\end{Prp}
\Proof The Lagrangian~$\L_\text{iso}$ is obtained from~\eqref{lagrangian_R3} by integrating
over~$\vartheta$. More precisely, 
\[ \L_\text{iso}(\tau, \tau')
= \frac1{16} \bigg( \int_{-1}^{\cos \vartheta_-} + \int_{\cos \vartheta_+}^1 \bigg)
\Big( \big( 1+\tau\tau'\cos\vartheta \big)^2-(\tau^2-1)(\tau'^2-1) \Big)\, d\cos \vartheta \:, \]
and a direct computation using~\eqref{eq:critical_angles} yields~\eqref{Liso}.
Evaluating this formula for the isotropic Lagrangian in the case~$\tau=\tau'=1$
gives the equality in~\eqref{eq:iso_measure_bound}.

It remains to prove the inequality in~\eqref{eq:iso_measure_bound}.
A simple way to verify this inequality is to plot the isotropic Lagrangian numerically,
showing that~$\tau=\tau'=1$ is indeed the unique minimum.
For the rigorous derivation of this inequality, by symmetry it suffices to consider the case~$\tau' \geq \tau$.
In the case~$\tau=\tau'$, the isotropic Lagrangian simplifies to
\beq \label{Ltt}
\L_\text{\rm{iso}}(\tau, \tau) = \frac{1}{4}-\frac{1}{12 \, \tau^{2}} \:,
\eeq
which clearly satisfies~\eqref{eq:iso_measure_bound}. Therefore, it remains to consider the case
\[ \tau' > \tau\:. \]
In this case, by direct computation we find that
\[ \frac{1}{\tau'}\: \frac{\partial\L_\text{\rm{iso}}(\tau, \tau)}{\partial \tau} - \frac{1}{\tau}\: \frac{\partial\L_\text{\rm{iso}}(\tau, \tau)}{\partial \tau'} = \frac{1}{4}\, \Big( \frac{1}{\tau^2} - \frac{1}{\tau'^2} \Big) \Big(
\sqrt{\tau^2-1}\, \sqrt{\tau'^2-1} - \tau \tau' \Big) < 0\:. \]
As a consequence, the function~$\L_\text{\rm{iso}}(\tau, \tau)$ is strictly decreasing along the
flow lines of the vector field~$(\tau'^{-1}, -\tau^{-1})$.
Starting from a point~$(\tau, \tau')$ with~$\tau'>\tau$, these flow lines describe
a circle which reaches a point in the parameter space with~$\tau=\tau'$.
This concludes the proof.
\QED

The lower bound for the Lagrangian~\eqref{eq:iso_measure_bound} immediately gives a corresponding
lower bound for the causal action,
\beq 
\Sact(\rho) \geq \frac{1}{6} \qquad \text{for isotropic measures~\eqref{spherical}} \:. \label{Sisolower}
\eeq
Moreover, the estimate~\eqref{eq:iso_measure_bound} shows that, varying within the class of isotropic measures,
the measure supported at~$\tau=1$,
\beq \label{rho1}
d\rho = \frac{1}{4 \pi}\: \delta(\tau-1) \: d\tau \:d\cos \vartheta\: d\varphi
\eeq
is the unique minimizer.
In the next section, we shall show by direct computation that non-isotropic measures
give rise to causal actions which are strictly smaller than~$1/6$.
In other words, isotropic measures are {\em{not}} minimizing.
Before entering this analysis, we mention a few previous related results.
For the measure~\eqref{rho1}, the causal action was already computed
in~\cite[Section~1.4]{continuum}. In~\cite[Sections~2 and~5]{support}, a detailed analysis was given
for the so-called {\em{causal variational principle on the sphere}}, where one restricts attention
to measures supported on a sphere of fixed radius~$\tau=\tau_0$.
In particular, in~\cite[Section~5.2]{support} the isotropic measure on the sphere was considered,
\[ 
d\rho = \frac{1}{4 \pi}\: \delta \big( \tau-\tau_0 \big) \: d\tau \:d\cos \vartheta\: d\varphi \:. \]
The corresponding causal action was computed, in agreement with
the formula~\eqref{Ltt} for the isotropic Lagrangian in the case~$\tau=\tau'$
(when comparing our formulas with those in~\cite{support}, one must keep in mind that
in the latter paper the normalization convention for the trace~\eqref{cnorm} is different by assuming
that this trace equals two; as a consequence, the causal Lagrangian
becomes larger by an overall factor of~$16$).

\subsection{The Two-Dimensional Discrete Dirac Sphere} \label{secdirac2d}
We now explain how to construct measures whose causal action is strictly smaller
than the lower bound~$1/6$ obtained for isotropic measures (see~\eqref{Sisolower}).
We again consider a normalized weighted counting measure (see~\eqref{weightcount} and~\eqref{cnorm}).
We choose the radial coordinates of all points equal to~$\tau_i=\tau$.
In other words, we distribute the points on a sphere of radius~$\tau$.
The resulting causal variational principle on the sphere was introduced in~\cite[Example~2.8]{continuum}
and analyzed further in~\cite{support}. 
This analysis revealed that the minimal causal action is attained for
configurations of points which are, as far as possible, equally distributed on the sphere.
More precisely, we choose the so-called {\em{Tammes distribution}} where
one maximizes the minimal distance~$\vartheta_m$ between any pair of points,
\[ \vartheta_m := \min_{i,j=1,\ldots, m} \arccos \big( \vec{x}_i \cdot \vec{x}_j \big) \]
(for details on the Tammes distribution see~\cite{saff+kuijlaars} or~\cite{sloane} and the references therein).
Moreover, one chooses~$\tau$ such that the opening angle of the light cone coincides with~$\vartheta_m$.
Thus, evaluating~\eqref{eq:critical_angles} in the case~$\tau=\tau'$, we choose
\beq \label{thetamin2}
\vartheta_{m}=\vartheta_- =  \arccos \!\left( 1-\frac{2}{\tau^2} \right) \:.
\eeq
This choice has the effect that all distinct points are spacelike separated, implying that it suffices to consider
the pair interaction of each point with itself, i.e.\
\[ \Sact(\rho) = \sum_{i=1}^m c_i^2 \: \L\big( (\tau, \vec{x}_i), (\tau, \vec{x}_i) \big) \:. \]
Finally, we choose all the weight factors equal to~$1/m$,
\[ \Sact(\rho) = \frac{1}{m^2} \sum_{i=1}^m \L\big( (\tau, \vec{x}_i), (\tau, \vec{x}_i) \big) \:. \]

Asymptotically for large~$m$, the Tammes distribution goes over to the densest sphere packing in~$\R^2$
with the radius of the spheres given by~$\vartheta_-/2$.
This makes it possible to compute the causal action asymptotically.

\begin{Prp} {\bf{(The two-dimensional discrete Dirac sphere)}} \label{propositionn1} Choosing the points on the sphere of radius~$\tau$ according to the Tammes distribution with~$\vartheta_m = \vartheta_-$, the parameter~$\tau$ and the causal action
have the following asymptotics for large~$m$,
\begin{align}
\tau &= \frac{\sqrt[4]{3}}{\sqrt{2 \pi}}\: \sqrt{m} + \O \Big( \frac{1}{\sqrt{m}} \Big) \label{tau_2dim_Dirac} \\
\Sact &= \frac{\sqrt{3}}{4 \pi} + \O \Big( \frac{1}{m} \Big) \:. \label{Sasy}
\end{align}
\end{Prp}
\Proof The packing density~$\delta_2$ of the densest sphere packing in~$\R^2$ is given by
(see~\cite[Section~1.5]{conway+sloane})
\[ \delta_2 = \frac{\pi}{6}\: \sqrt{3} \:. \]
As a consequence, the number~$m$ of circles of radius~$\vartheta_-/2$
which can be distributed on the unit sphere~$S^2$ is given asymptotically by
\begin{align}
m &= \frac{\delta_2\: \mu(S^2)}{\mu \big( B^2_{\vartheta_-/2} \big)} \: \Big( 1 + \O(\vartheta_-) \Big) 
= \frac{\pi}{6}\: \sqrt{3}\: \frac{4 \pi}{\pi\, (\vartheta_-/2)^2} + \O\Big( \frac{1}{\vartheta_-} \Big) \notag \\
&= \frac{8 \pi}{\sqrt{3}}\: \frac{1}{\vartheta_-^2} + \O\Big( \frac{1}{\vartheta_-} \Big)\:. \label{masy1}
\end{align}
Next, using~\eqref{thetamin2},
\beq \label{tauasy}
1-\frac{2}{\tau^2} = \cos \vartheta_- = 1 -\frac{\vartheta_-^2}{2} + \O\big( \vartheta_-^4 \big)
\eeq
and thus
\[ \vartheta_- = \frac{2}{\tau} + \O\Big( \frac{1}{\tau^2} \Big) \:. \]
Using this expansion in~\eqref{masy1}, we conclude that
\[ \tau^2 = \frac{\sqrt{3}}{2 \pi}\: m + \O \big( m^0 \big) \:. \]

For this configuration, we only get contributions to the causal action if~$x_i=x_j$.
Thus
\[ \Sact = \frac{1}{m^2} \sum_{i=1}^m \L\big( F_\tau(\vec x_i), F_\tau(\vec x_i) \big) = \frac{1}{m}\:
\frac{\tau^2}{8}\: 2\cdot2 = \frac{\tau^2}{2m} 
=  \frac{\sqrt{3}}{4 \pi} + \O \Big( \frac{1}{m} \Big) \:, \]
concluding the proof.
\QED

Comparing the asymptotics~\eqref{Sasy} with the estimate~\eqref{Sisolower} and using that
\[  0.138 \approx \frac{\sqrt{3}}{4 \pi} \;<\; \frac{1}{6} \approx 0.167 \:, \]
we conclude that, for large~$m$, the absolute minimizer of the causal action principle cannot be isotropic.
This result generalizes previous finding for the variational principle on the sphere
in~\cite{support}; see also the finer analytic results in~\cite{sphere}.

Another point of interest, which will also be important in our numerical analysis,
is that, according to~\eqref{tau_2dim_Dirac}, the parameter~$\tau$ tends to infinity as~$m \rightarrow \infty$.
In other words, the larger we choose~$m$, the larger the radius of the discrete Dirac sphere of minimal
action gets. This finding corresponds to a {\em{non-compactness}} of the causal action principle
as first observed in~\cite[Section~2.2]{continuum}, which can be stated mathematically that
a minimizing sequence of the causal action may diverge in the sense that
the support of the measures tends to infinity.
In order to gain compactness, in~\cite{continuum, cfs} the {\em{boundedness constraint}}~\eqref{Tdef}
was introduced. In our setting, the necessity of the boundedness constraint corresponds to the fact
that, analyzing sequences of minimizing measure~$(\rho_\ell)_{\ell \in \N}$ asymptotically as~$m \rightarrow \infty$,
these sequences of measures will typically diverge, while the functional~$\T$ 
in~\eqref{Tdef} will tend to infinity.

With this in mind, we now compute the functional~$\T$ for the discrete Dirac sphere.
\begin{Prp}\label{propositionn1T} For the measure describing the two-dimensional discrete Dirac sphere
in Proposition~\ref{propositionn1}, the functional~$\T$ in~\eqref{Tdef} has the asymptotics
\[ \T
=  \frac{3}{16\, \pi^2}\:m^2 +\O(m) \:. \]
\end{Prp}
\Proof
 Again using the bijection~\eqref{eq:R3DiracMap} in order to parametrize the points of~$\F$
by~$(\tau, \vec{x})$, functional~$\T$ becomes
\[ \T(\rho)=\int d \rho(\tau,\vec x)\int d\rho(\tau',\vec y) \:\big( |\lambda_1|+|\lambda_2| \big)^2 \:, \]
where  $\lambda_{1,2}$ are the eigenvalues of the product $F_\tau(\vec x)F_{\tau'}(\vec x')$ in \eqref{evals_product}.
If the points $(\tau,\vec x)$ and $(\tau',\vec x')$ are {\em{spacelike}} separated, these eigenvalues
form a complex conjugate pair (because the square root is zero or imaginary). Hence, in this case,
\[\big( |\lambda_1|+|\lambda_2| \big)^2 = 4 \,|\lambda_1|^2 = 4\, \lambda_1 \lambda_2
\overset{\eqref{prodsign}}{=}
=\frac{1}{4}\: (\tau^2-1)(\tau'^2-1) \:. \]
On the other hand, if the points are {\em{timelike}} separated, both eigenvalues are real and have the same sign.
Therefore,
\[ \big( |\lambda_1|+|\lambda_2| \big)^2 = \big( \lambda_1 + \lambda_2 \big)^2
= \Big( \tr\big( F_\tau(\vec{x}) \, F_\tau(\vec{y}) \big) \Big)^2
\overset{\eqref{FF}}{=}
\frac{1}{4}\: \big(1+\tau\tau'\cos \vartheta \big)^2 \:. \]

As in Proposition~\ref{propositionn1}, we now choose the weights~$c_i=1/m$, and the~$\vec x_i$
such that all distinct points are spacelike separated.
Then every single point contributes $(1+\tau^2)^2/4$ to the~$\T$, whereas each (ordered) pair of different points contributes $(\tau^2-1)^2/4$. Thus
\[ 
\T(\rho) = \frac1{m^2}\: \Big( m \:\frac{(1+\tau^2)^2}4 + m \,(m-1) \:\frac{(\tau^2-1)^2}4 \Big)
=\frac{(\tau^2-1)^2}{4} +\frac{\tau^2}{m} \:, \]
which shows that, up to a term of order $\O(m^{-1})$, the boundedness constraint is determined
by pairs of points with spacelike separation. Employing the asymptotics for~$\tau$ in~\eqref{tau_2dim_Dirac}
gives the result.
\QED

\subsection{Causally Trivial Measures as Local Minimizers} \label{sectivcausal}
In the example of the two-dimensional discrete Dirac sphere
(see Proposition~\ref{propositionn1}), all distinct spacetime points are spacelike separated, i.e.\
\[ \L(x_i, x_j)=0 \qquad \text{for all~$i \neq j$} \:. \]
We refer to a measure~$\rho$ with this property as being {\em{causally trivial}}.
We conjecture that if the dimension~$f$ of the Hilbert space is sufficiently small, then every
minimizing measure should be causally trivial. If~$f$ is sufficiently large, however,
the causal structure of minimizing measures should become non-trivial.
As a first step towards testing this conjecture, in this section we conclude the analysis in the case~$f=2$ by explaining
a mechanism which shows why causally trivial measures give rise to local minima of the causal action.

Again using the identification~\eqref{eq:R3DiracMap}, we write the measure~\eqref{weightdelta} as
\[ \rho=\sum_{i=1}^m c_i \,\delta_{F_{\tau_i}(\vec x_i)} \:. \]
If this measure is causally trivial, in the causal action only the summands with~$i=j$ contribute,
\[ 
\Sact(\rho) = \sum_{i=1}^m c_i^2 \:\L \big( (\tau_i, \vec x_i), (\tau_i, \vec x_i) \big)
=\sum_{i=1}^m c_i^2 \:\frac{\tau_i^2}2 \:, \]
where in the last step we applied~\eqref{Ldiag}.
The weights~$c_i$ can be determined by mi\-ni\-mizing the causal action under variations under the
constraints~\eqref{cnorm}. Using the Lagrange multiplier method, one readily finds the optimal weights
\beq \label{civals}
c_i = \frac{1}{\tau_i^2}\: \bigg( \sum_{j=1}^m \frac{1}{\tau_j^2} \bigg)^{-1} \:.
\eeq
As a consequence, the causal action simplifies to
\[ \Sact(\rho) = \frac{1}{2} \bigg( \sum_{i=1}^m \frac{1}{\tau_i^2} \bigg)^{-1} \:. \]

We shall now give a sufficient condition for such measures to be local minimizers of the causal action.
To this end, we introduce the following notion.
\begin{Def}
A point $(\tau_i,\vec x_i)$ in spacetime~$M:= \supp \rho$ of a causally trivial measure~$\rho$
is said to be {\bf{angular lightlike restricted}} if there exists an open neighborhood~$U\subset S^2$ of~$\vec x_i$ such that for all $\vec y\in U$ the point $(\tau_i,\vec y)$ is timelike separated from at least one  point in $M \setminus \{ (\tau_i,\vec x_i)\}$.
\end{Def} \noindent
Intuitively speaking,
angular lightlike restriction means that the angular coordinate $\vec x_i\in S^2$ of a point $(\tau_i,\vec x_i)\in \F$ cannot be varied without the point entering the light cone of another spacetime point.

\begin{Prp} Let~$\rho$ be a causally trivial measure with optimal weights~\eqref{civals}.
Assume that every spacetime point~$(\tau_i,\vec x_i)$ in $\supp \rho$ is angular lightlike restricted and 
lies on the boundary of the light cone for at least $k\geq 3$ points. Moreover, assume that the radial coordinates of all points are at least
\[ \tau_i\geq\sqrt{\frac{k}{k-2}} \:. \]
Then~$\rho$ is a local minimizer of the causal action principle.
\end{Prp}
\Proof
We need to show that the action~$\Sact(\rho)$ cannot be decreased by a small
change in the free parameters, namely the weights, the angular coordinates or the radial coordinates.
The weights are already chosen optimally by~\eqref{civals}. Moreover, since all points are angular restricted,
any non-trivial variation of the angular coordinates~$x_i$
gives rise to a strictly positive first variation of~$\Sact(\rho)$.
Hence it remains to show that any small change of the radial coordinates~$\tau_i$ increases $\Sact(\rho)$.

The contribution from a pair of points to the total action is given in \eqref{lagrangian_R3}. 
Differentiating the self-contribution in~\eqref{Ldiag} gives
\[ \partial_{\tau_i} \L(F_{\tau_i}(\vec x_i),F_{\tau_i}(\vec x_i))= \tau_i \:. \]
Thus the self-contribution becomes smaller when the radial coordinate of a point is decreased.
We next consider a pair of points~$(\tau,\vec x)$ and~$(\tau',\vec x')$ which are
on the boundary of the light cone such that the angle $\vartheta$ between $\vec x$ and $\vec x'$ is smaller than~$\frac{\pi}{2}$, i.e.,
according to~\eqref{eq:critical_angles},
\[ \cos\vartheta=\cos\vartheta_-=\frac1{\tau \tau'}\left(\sqrt{(\tau^2-1)(\tau'^2-1)}-1\right) .  \]
In this case, the contribution~$\L(F_{\tau}(\vec x),F_{\tau'}(\vec x'))$ to the causal action vanishes.
This term remains zero if either of the radial coordinates are {\em{increased}}. Hence the one-sided derivative, corresponding to an increase of the radial coordinate, vanishes:
\[ \frac\partial{\partial \epsilon} \bigg|_{\epsilon\geq0} \L \big( (\tau+ \epsilon, \vec x), (\tau', \vec x') \big)=0 \:. \]
If, however, the radial coordinate of one point is {\em{de}}creased, we obtain
\begin{align*}
\frac\partial{\partial \epsilon} \bigg|_{\epsilon\geq0} \L \big( (\tau-\epsilon, \vec x), (\tau', \vec x') \big)
&= - \frac1{8} \frac\partial{\partial \tau} \Big( (1+\tau\tau'\cos\vartheta_-)^2-(\tau^2-1)(\tau'^2-1) \Big) \\
&= \frac{\sqrt{\tau'^2-1}}{4 \tau}\: \Big(\sqrt{\tau'^2-1}+\sqrt{\tau^2-1} \Big) \:.
\end{align*}

Now consider an arbitrary spacetime point $(\tau_i,\vec x_i)\in\supp \rho$
and let $\{ (\tau_j,\vec x_j) \,|\, j\in J\}$ be the set of points which lie on the light cone centered at~$(\tau_i,\vec x_i)$. (By assumption we have $\# J\geq k$, i.e.\ $J$ contains at least $k$ elements.)
Then the one-sided derivative of the total action when decreasing the radial coordinate $\tau_i$ is given by
\begin{align*}
\frac\partial{\partial \epsilon} &\bigg|_{\epsilon\geq0} c_i^2 \:
\L \big( (\tau_i-\epsilon, \vec x_i), (\tau_i-\epsilon, \vec x_i) \big) + 2 c_i \sum_{j \in J} c_j \:
\L \big( (\tau_i-\epsilon, \vec x_i), (\tau_j, \vec x_j) \big) \\
&= -c_i^2\: \tau_i + 2 c_i \sum_{j \in J} c_j \: \frac{\sqrt{\tau_j^2-1}}{4 \tau_i}\: \bigg(
\sqrt{\tau_j^2-1}+\sqrt{\tau_i^2-1} \bigg) \\
&= \bigg( \sum_{k=1}^m \frac{1}{\tau_k^2} \bigg)^{-2} \,
\bigg\{ -\frac{1}{\tau_i^3} + 2 \sum_{j \in J} \frac{1}{\tau_i^2 \tau_j^2} \: \frac{\sqrt{\tau_j^2-1}}{4 \tau_i}\: \bigg(
\sqrt{\tau_j^2-1}+\sqrt{\tau_i^2-1} \bigg) \bigg\} \\
&> \bigg( \sum_{k=1}^m \frac{1}{\tau_k^2} \bigg)^{-2} \,\frac{1}{\tau_i^3}\,
\bigg\{ -1 + \sum_{j \in J} \frac{\sqrt{\tau_j^2-1}}{2 \tau_j^2}\;
\sqrt{\tau_j^2-1} \bigg\} \\
&= \bigg( \sum_{k=1}^m \frac{1}{\tau_k^2} \bigg)^{-2} \,\frac{1}{\tau_i^3}\,
\bigg\{ -1 + \sum_{j \in J} \Big( \frac{1}{2}  - \frac{1}{2 \tau_j^2} \Big) \bigg\} \\
&\geq \bigg( \sum_{k=1}^m \frac{1}{\tau_k^2} \bigg)^{-2} \,\frac1{\tau_i^3}\,
\bigg\{- 1 + k \Big(\frac12-\frac{k-2}{2 k }\Big) \bigg\} = 0 \:.
\end{align*}
Hence the variation is strictly positive, concluding the proof.
\QED

\section{The Four-Dimensional Discrete Dirac Sphere} \label{secdirac4d}
In the previous section, we considered the simplest interesting case~$n=1$ and~$f=2$.
For physically interesting models one should choose the dimension~$f$ of the Hilbert space~$\H$
to be very large. Moreover, in order to model Dirac spinors, one should choose the spin dimension~$n=2$.
The resulting higher-dimensional causal fermion systems are much more difficult to analyze,
mainly because the dimension of the space~$\F$ of linear operators gets large
(more precisely, the operators with~$n$ positive and~$n$ negative eigenvalues and prescribed trace form a
manifold of dimension~$4n f - 4n^2 -1$; see~\cite[Theorem~3.2]{gaugefix}).
In particular, one no longer has the simple representation of the operators in~$\F$ in terms of Pauli
matrices~\eqref{eq:R3DiracMap}. For this reason, there are hardly any analytic results,
making it necessary to explore the situation numerically.

However, there are higher-dimensional analogs of the discrete Dirac spheres of Proposition~\ref{propositionn1}.
In the case~$n=2$ and~$f=4$, the resulting causal fermion systems will serve as test cases
for the numerics. We conjecture that, asymptotically for large~$m$, minimizers of the 
causal action principle should be close to these discrete Dirac spheres.
Compared to the two-dimensional case, there is the additional feature that the causal action
tends to zero if~$m \rightarrow \infty$ (see Proposition~\ref{propositionn2}).
The resulting sequence of measures again diverges,
in agreement with our discussion of the boundedness constraint at the end of Section~\ref{secdirac2d}.
The new feature is that there are divergent minimizing sequences whose causal action tends to zero.

We now introduce the four-dimensional discrete Dirac sphere in detail,
based on a related continuous system introduced in~\cite[Example~2.9]{continuum}
(the so-called three-dimensional Dirac sphere).
Let~$\H=\C^4$. We introduce five $4 \times 4$-matrices acting on~$\H$ by
\[ \gamma^\alpha = \begin{pmatrix} \sigma^\alpha & 0 \\ 0 & -\sigma^\alpha \end{pmatrix},
\quad \alpha=1,2,3  \:, \qquad \gamma^4 = \begin{pmatrix} 0 & i\,\1 \\ -i\,\1 & 0 \end{pmatrix} \qquad
\text{and} \qquad \gamma^5 = \begin{pmatrix} 0 & \1 \\ \1 & 0 \end{pmatrix} \]
(where the~$\sigma^\alpha$ are again the Pauli matrices~\eqref{pauli}).
Given a parameter~$\tau>1$, we consider the following mapping from
the sphere~$S^4 \subset \R^5$ to the linear operators on~$\H$,
\[ F \::\: S^4 \rightarrow \Lin(\H) \:,\qquad F(x) = \frac{1}{4} \bigg( \sum_{i=1}^5 \tau\: x^i \gamma^i + \1
\bigg) \:. \]
Choosing~$m$ points~$x_1, \ldots, x_m \in S^4$ on the sphere, we again introduce~$\rho$
as the normalized weighted counting measure~\eqref{weightcount}.

Let us verify that the matrix~$F(x)$ lies in~$\F$: First, this matrix is obviously symmetric and has trace one.
In order to compute its eigenvalues, it is most convenient to make use of the fact that the matrices~$\gamma^j$ are the Dirac matrices of Euclidean~$\R^5$, satisfying the anti-commutation relations
\[ \{\gamma^i, \gamma^j\} = 2 \delta^{ij}\:\1\qquad (i,j=1,\ldots, 4)\:. \]
As a consequence,
\begin{align*}
4 F(x) -\1 &= \sum_{i=1}^5 \tau\: x^i \gamma^i \\
\big(4 F(x) -\1\big)^2 &= \sum_{i,j=1}^5 \tau^2\: x^i\,x^j \gamma^i \gamma^j
= \frac{\tau^2}{2} \sum_{i,j=1}^5 x^i\,x^j \big\{\gamma^i, \gamma^j \big\} \\
&= \frac{\tau^2}{2} \sum_{i,j=1}^5 x^i\,x^j \:2\, \delta_{ij} \:\1 = \tau^2\, \1 \:.
\end{align*}
Hence the matrix~$F(x)$ satisfies the polynomial equation
\[ \big(4 F(x) -\1\big)^2 = \tau^2\, \1 \:. \]
We conclude that~$F(x)$ has the eigenvalues
\[ 
\nu_\pm = \frac{1}{4}\: \big( 1 \pm \tau \big) \:. \]
Moreover, since the matrix~$4F(x)-\1$ is trace-free, each eigenvalue must appear with multiplicity two.
We conclude that~$F(x) \in \F$ if we choose the spin dimension~$n=2$.

We next compute the Lagrangian.
\begin{Lemma}
Denoting the angle between the vectors~$x,y \in \R^5$ by~$\vartheta$,
\begin{align*}
\L(F(x), F(y) \big)
&= \frac{\tau^2}{64}\: (1+\cos \vartheta) \left( 2 - \tau^2 \:(1-\cos \vartheta) \right) \:
\Theta(\vartheta_{\max}-\vartheta) \:,
\end{align*}
where
\beq \label{thetamax4}
\vartheta_{\max} := \arccos \!\left( 1-\frac{2}{\tau^2} \right) \:.
\eeq
\end{Lemma}
\Proof Let us compute the eigenvalues of the operator product~$F(x)\, F(y)$.
Again, it is most convenient to make use of the Clifford relations. First,
\begin{align}
16 \,F(x) \:F(y) &= \Big( \sum_{i=1}^5 \tau\: x^i \gamma^i + \1 \Big) \Big( \sum_{j=1}^5 \tau\: y^j \gamma^j + \1 \Big) \notag \\
&= \big(1 + \tau^2 \:\la x,y \ra \big) \1 + \tau \sum_{k=1}^5 (x^k+y^k) \gamma^k
+ \frac{\tau^2}{2} \:\sum_{i,j=1}^5 x^i y^j \, \big[\gamma^i, \gamma^j\big] \:. \label{trrel}
\end{align}
Using that
\[ \gamma^i\, \big[\gamma^i, \gamma^j \big] = -\big[\gamma^i, \gamma^j \big]\: \gamma^i \:, \]
we conclude that
\begin{align*}
\Big(16\,F(x) \:F(y) - \big(1 + \tau^2 \:\la x,y \ra \big) \1 \Big)^2
= \tau^2 \sum_{k=1}^5 \big(x^k+y^k \big)^2 \1 + 
\bigg( \frac{\tau^2}{2} \sum_{i,j=1}^5 x^i y^j \, \big[\gamma^i, \gamma^j\big] \bigg)^2 \:.
\end{align*}
This can be simplified with the help of the relations
\begin{align}
\sum_{i=1}^5 (x^i+y^i)^2 &= 2 + 2 \,\la x,y \ra \\
\bigg( \sum_{i,j=1}^5 x^i y^j \, \big[\gamma^i, \gamma^j\big] \bigg)^2
&= -4 \sin^2 \vartheta \,\1 = -4\,\big( 1-\la x,y \ra^2 \big)\, \1\:, \label{blin}
\end{align}
where~$\vartheta$ is the angle between the vectors~$x,y \in \R^5$.
The relation~\eqref{blin} can be verified in detail as follows. The rotational symmetry of the
Euclidean Dirac operator on~$\R^5$ means that for every rotation~$O \in \text{SO}(5)$ there is
a unitary operator~$U \in \text{SU}(4)$ such that
\[ O^i_j\, \gamma^j = U \gamma^i U^{-1} \]
(more details can be found in the three-dimensional case in~\cite[Section~1]{snygg}
and in four dimensions in~\cite[Section~2.1]{snygg}, \cite[Section~2.2]{bjorken} 
or~\cite[eq.~(3.29)]{peskin+schroeder}; the five-dimensional case considered here is similar).
Making use of this rotational symmetry, we can arrange that the vector~$x$ is the basis vector~$e_1$
and that~$y = \cos \vartheta\, e_1 + \sin \vartheta\, e_2$. As a consequence,
\begin{align*}
\sum_{i,j=1}^5 x^i y^j \, \big[\gamma^i, \gamma^j \big] &= \sin \vartheta\: [\gamma^1, \gamma^2]
= 2\, \sin \vartheta\: \gamma^1 \gamma^2 \\
\bigg( \sum_{i,j=1}^5 x^i y^j \, \big[\gamma^i, \gamma^j\big] \bigg)^2 &= 
4\, \sin^2 \vartheta\: \gamma^1 \gamma^2 \gamma^1 \gamma^2 \:,
\end{align*}
and applying the anti-commutation relations gives~\eqref{blin}.

Combining the above equations, we conclude that the product~$F(x) \:F(y)$ satisfies the polynomial equation
\begin{align*}
\Big(16\, F(x) \:F(y) - \big(1 + \tau^2 \:\la x,y \ra \big) \1 \Big)^2
&= 2 \,\tau^2 \big(1+ \la x,y \ra \big)\1 - \tau^4 \:\big(1-\la x,y \ra^2 \big) \1 \\
&= \tau^2 \:\Big(1+ \la x,y \ra \Big) \Big( 2 - \tau^2 \:\big(1-\la x,y \ra \big) \Big)\1 \:.
\end{align*}
Taking the square root, the zeros of this polynomial are computed by
\beq \label{leigen4}
\lambda_{1\!/\!2} = \frac{1}{16} \Big( 1+\tau^2 \: \la x,y \ra \pm
 \tau \,\sqrt{1+\la x,y \ra}\: \sqrt{2 - \tau^2 \:(1-\la x,y \ra)} \Big) \:.
\eeq
Moreover, taking the trace of~\eqref{trrel}, one finds
\[ \tr \big( 16\,F(x) \:F(y) \big) = 4\,\big(1 + \tau^2 \:\la x,y \ra \big) \:. \]
This implies that each eigenvalue in~\eqref{leigen4} has algebraic multiplicity two.

After these preparations, we can compute the Lagrangian.
If~$\vartheta$ is sufficiently small, then the term~$(1-\la x,y \ra)$ in~\eqref{leigen4} is close to zero, and thus the
arguments of the square roots are all positive. However, if~$\vartheta$ becomes so large that~$\vartheta
\geq \vartheta_{\max}$, then the argument of the last square root in~\eqref{leigen4} becomes negative, so that
the eigenvalues~$\lambda_{1\!/\!2}$ form a complex conjugate pair. Moreover, a short calculation shows that
\beq \label{lprod4}
\lambda_1 \lambda_2 = \frac{1}{256}\: (1+\tau)^2 (1-\tau)^2 > 0 \:,
\eeq
implying that if the~$\lambda_{1\!/\!2}$ are both real, then they have the same sign.
Using this information in~\eqref{Lagrange}, the Lagrangian simplifies to
\begin{align*}
\L(F(x), F(y) \big) &= \frac{1}{8} \sum_{i,j=1}^{4} \Big( \big|\lambda^{xy}_i \big| - \big|\lambda^{xy}_j \big| \Big)^2 
= \frac{1}{2} \sum_{i,j=1}^{2} \Big( \big|\lambda_i \big| - \big|\lambda_j \big| \Big)^2 \\
&= \frac{1}{2}\:\Theta(\vartheta_{\max}-\vartheta)
\sum_{i,j=1}^{2} \Big( \lambda_i -\lambda_j \Big)^2 
= \Theta(\vartheta_{\max}-\vartheta) \big( \lambda_1 -\lambda_2 \big)^2 \\
&= \frac{4 \,\tau^2}{256}\: (1+\cos \vartheta) \left( 2 - \tau^2 \:(1-\cos \vartheta) \right) \:
\Theta(\vartheta_{\max}-\vartheta) \:.
\end{align*}
This concludes the proof.
\QED

\begin{Prp}\label{propositionn2} Choosing the points on~$S^4$ according to the Tammes distribution
on the four-sphere with~$\vartheta_m = \vartheta_{\max}$ and
\beq \label{tauval4}
\tau = \frac{3^\frac{1}{4}}{\sqrt{\pi}}\: \sqrt[4]{m} + \O \big( m^{-\frac{3}{4}} \big) \:,
\eeq
the causal action has the asymptotics
\[ \Sact = \frac{\sqrt{3}}{16\, \pi} \frac{1}{\sqrt{m}} + \O \big( m^{-\frac{3}{2}} \big) \:. \]
\end{Prp}
\Proof The packing density~$\delta_4$ of the densest sphere packing in~$\R^4$ is given by
(see~\cite[Section~1.5]{conway+sloane})
\[ \delta_4 = \frac{\pi^2}{16} \:. \]
Moreover, the volume~$\mu(S^4)$ of the unit $4$-sphere
and the volume~$\mu(B^4_r)$ of a $4$-ball are
\[ \mu(S^4) = \frac{8}{3}\: \pi^2 \qquad \text{and} \qquad \mu(B^4_r) = \frac{\pi^2}{2}\: r^4 \:. \]
As a consequence, the number~$m$ of $4$-balls of radius~$\vartheta_{\max}/2$
which can be distributed on the unit sphere~$S^4$ is given asymptotically by
\begin{align}
m &= \frac{\delta_4\: \mu(S^4)}{\mu \big( B^4_{\vartheta_{\max}/2} \big)} \: \Big( 1 + \O(\vartheta_{\max}) \Big) 
= \frac{\pi^2}{16} \: \frac{8}{3}\: \pi^2 \frac{2}{\pi^2} \: \frac{1}{(\vartheta_{\max}/2)^4}
+ \O\Big( \frac{1}{\vartheta^3_{\max}} \Big) \notag \\
&= \frac{16\, \pi^2}{3}\: \frac{1}{\vartheta_{\max}^4} + \O\Big( \frac{1}{\vartheta^3_{\max}} \Big) \:. \label{masy2}
\end{align}
Next, using~\eqref{thetamax4}, exactly as in~\eqref{tauasy} we obtain
\[ \vartheta_{\max} = \frac{2}{\tau} + \O\Big( \frac{1}{\tau^2} \Big) \:. \]
Using this expansion in~\eqref{masy2}, we conclude that
\[ \tau^4 = \frac{3}{\pi^2}\: m + \O \big( m^0 \big) \:. \]

For this configuration, we only get contributions to the causal action if~$x=y$. Thus
\[ \Sact = \frac{1}{m^2} \sum_{i=1}^m \L\big( F(x_i), F(x_i) \big) = \frac{1}{m}\:
\frac{\tau^2}{64}\: 2\,2 = \frac{1}{16}\: \frac{\sqrt{3}}{\pi}\: \frac{1}{\sqrt{m}} + \O \big( m^{-\frac{3}{2}} \big) \:. \]
This gives the result.
\QED

Similar as in Proposition~\ref{propositionn1T}, we finally compute the functional~$\T$.
\begin{Prp}\label{prpT4} Choosing the points on~$S^4$ according to the Tammes distribution
on the four-sphere with~$\vartheta_m = \vartheta_{\max}$ and~$\tau$ according to~\eqref{tauval4},
the functional~$\T$ in~\eqref{Tdef} has the asymptotics
\beq \label{Tform}
\T =  \frac{1}{16}\: (1+\tau)^2 (1-\tau)^2 + \O \Big( \frac{1}{m} \Big) \:.
\eeq
\end{Prp}
\Proof Exactly as in the proof of Proposition~\ref{propositionn1T}, the leading contribution to~$\T$
comes from pairs of points with spacelike separation. In this case, the eigenvalues of the
operator product~$F(x)\, F(y)$ form a complex conjugate pair and are both two-fold degenerate.
Moreover, using~\eqref{leigen4} and~\eqref{lprod4}, we obtain
\[ |\lambda_1|^2 = |\lambda_2|^2 = \lambda_1 \lambda_2 = 
\frac{1}{256}\: (1+\tau)^2 (1-\tau)^2 \:, \]
and thus
\[ \bigg( \sum_{i=1}^{2n} \big|\lambda^{xy}_i \big| \bigg)^2 =  \frac{1}{16}\: (1+\tau)^2 (1-\tau)^2 \:. \]
Integrating over~$x$ and~$y$ gives the leading term in~\eqref{Tform}.
The pairs of points with timelike separation, however, contribute to the order~$1/m$ and can therefore be
absorbed into the error term.
\QED

\section{Causal Cones in Spin Dimension One} \label{seccausalgen}
In Section~\ref{seclagcausal} we showed that in the case~$n=1$ and~$f=2$, the causal relations
give rise to cones (see Figure~\ref{figlightcone}).
This geometric picture extends to a general dimension~$f$ of the Hilbert space, as we now explain.
While it is not possible to represent the causal relations between all points in $\F$ simultaneously, the relation between one fixed point and all other points can be represented geometrically again with a double cone
(see Figure~\ref{figlightcone2} below).
We again choose the spin dimension~$n=1$, but now with arbitrary~$f \geq 2$.
We denote the fixed reference point by~$x\in\F$, and its non-trivial eigenvalues
by~$\nu_\pm=\frac12(1\pm\tau)$ with $\tau\geq1$, as above. 
Furthermore, we denote the orthogonal projection to the two-dimensional image of $x$ by
\[ \pi_x \::\: \H \rightarrow x(\H) \subset \H \:. \]
Let $y\in\F$ be arbitrary. The causal relations between~$x$ and~$y$ are determined
by the eigenvalues of the product $xy$. These eigenvalues are equal to the eigenvalues of
the operator product~$x(\pi_x y \pi_x)$. Hence, in order to analyze the causal relation between $x$ and $y$ it suffices to investigate the projected operator $\pi_x y \pi_x$. We first derive inequalities for its eigenvalues.

\begin{Prp} \label{prpproject}
Let $\nu'_+>0$ and $\nu'_-=(1-\nu'_+)\leq0$ denote the non-trivial eigenvalues of $y$. Then the
eigenvalues~$\nu'_1,\nu'_2$ of the operator~$\pi_xy\pi_x$, ordered by size~$\nu'_1 \leq \nu'_2$, satisfy the inequalities
\begin{align}\label{eq:restriction_projected_evs}
\nu'_- \leq \nu'_1 \leq 0 \leq \nu'_2 \leq \nu'_+ \:.
\end{align}
Furthermore, the trace of the projected operator is bounded by
\[ \nu'_-\leq \tr \big( \pi_xy\pi_x \big) \leq \nu'_+ \:. \]
\end{Prp}
\Proof The lemma follows readily from the min-max principle (see for example~\cite[Section~XIII.1]{reed+simon4}).
Indeed, the inequality~$\nu'_2 \leq \nu'_+$ follows from the estimate
\[ \nu'_+ = \sup_{0 \neq u \in \H} \frac{\la u | y u \ra_\H}{\la u | u \ra_\H}
\geq \sup_{0 \neq u \in x(\H)} \frac{\la u | y u \ra_\H}{\la u | u \ra_\H}
= \sup_{0 \neq u \H} \frac{\la u \,|\, \pi_x y \pi_x\, u \ra_\H}{\la u | u \ra_\H}
= \nu'_2 \:. \]
The proof of the inequality~$\nu'_- \leq \nu'_2$ is analogous.

In order to show that one eigenvalue is non-negative and the other is non-positive,
we proceed indirectly. Assume conversely that both eigenvalues are, for example, strictly positive,
\[ \nu'_1, \nu'_2 > 0 \:. \]
Then the  min-max principle implies that the operator~$y$
has at least two strictly positive eigenvalues, which is a contradiction to $y\in\F$.

Finally, the estimate of the trace follows immediately from the identity~$\tr \pi_xy\pi_x=\nu'_1+\nu'_2$
by applying~\eqref{eq:restriction_projected_evs}.
\QED

We choose an orthonormal eigenvector basis of~$x$ of the subspace~$x(\H)$. In this basis,
\beq \label{xbasisrep}
x|_{x(\H)} = \frac{1}{2}\: \big( \1+ \tau \: \sigma^3 \big) \:.
\eeq
Similarly to~\eqref{eq:R3DiracMap}, the operator $\pi_xy\pi_x$ can be represented in this basis
in terms of four real coordinates~$(y_0,y_1,y_2,y_3)$ by the matrix
\begin{align}\label{eq:matrix_PyP}
\pi_x \, y\, \pi_x \big|_{x(\H)} = \frac{1}{2}\: \big(y_0\1+\sum_{i=1}^3 y_i\sigma_i \big)
=\frac12\begin{pmatrix}y_0+y_3 & y_1-\ii y_2\\ y_1+\ii y_2 & y_0 - y_3\end{pmatrix}.
\end{align}
This matrix also corresponds to the first diagonal $2\times2$-block of the $f\times f$-dimensional matrix representation of $y$ on $\H$ with respect to the eigenvector basis of~$x$ (in which the two non-trivial eigenvectors appear first). The eigenvalues of this matrix are computed by
\[ \nu'_{1\!/\!2} = \frac{1}{2} \big( y_0 \pm |\vec{y}| \big) \]
(with~$|\vec{y}|^2 = y_1^2+y_2^2+y_3^2$). In order to get agreement with the formula~\eqref{nupmform}
in Section~\ref{seclagcausal}, we denote the eigenvalues of~$y$ by
\[ \nu'_{\pm} =  \frac{1}{2} \:\big( 1 \pm \tau' \big) \:. \]
Then the inequalities~\eqref{eq:restriction_projected_evs} imply
the side conditions
\begin{align} \label{eq:sideconditions}
0\leq y_0 + |\vec{y}| \leq 1+\tau' \qquad \text{and} \qquad 1-\tau'\leq y_0-|\vec{y}| \leq 0
\end{align}
on the coordinates $(y_0,y_1,y_2,y_3)$.
This means that the allowed values of the coordinates~$(y_0, \vec{y})$
representing the operator~$\pi_xy\pi_x$, for any $y\in \F$ are restricted to lie inside the
shaded region shown on the right of Figure~\ref{figlightcone2}. In particular, the allowed region lies
outside the double cone defined by $|\vec{y}| \geq |y_0|$.
In the following, we will see that this cone generalizes the unit circle around the origin in
Figure~\ref{figlightcone}, which corresponds to the hyperplane $y_0=1$.

The point $x$ can be represented on~$x(\H)$ again through~\eqref{eq:R3DiracMap} by $F_\tau(\vec x)$. Using this representation, the eigenvalues of the matrix~$x \,\pi_xy\pi_x$ are computed by
\beq \label{eq54}
\lambda_{1,2}= \frac14 \:\bigg( y_0+y_3 \tau\pm\sqrt{(y_1^2+y_2^2)(1-\tau^2)+(y_3+y_0\tau)^2} \bigg) \:.
\eeq
It follows that the point $y$ is spacelike separated from $x$ if and only if
\beq \label{relcausal}
(\tau^2-1)(y_1^2+y_2^2)>(y_3+y_0 \tau)^2.
\eeq
In every hyperplane of constant~$y_0$, this equation defines a double cone with apex at the
point~$(y_0,y_1=y_2=0,y_3=-y_0 \tau)$ and opening angle~$2\theta$
determined by the equation~$\sin\theta=1/\tau$. This means geometrically
that the cone is tangent to the sphere of
radius~$|y_0|$ around the origin.
The resulting cone in the hyperplane of constant~$y_0$ is depicted on the left of Figure~\ref{figlightcone2}.
\begin{figure}
\psscalebox{1.0 1.0} 
{
\begin{pspicture}(0,26.17432)(12.719405,33.555927)
\definecolor{colour0}{rgb}{0.9019608,0.9019608,0.9019608}
\definecolor{colour1}{rgb}{0.7019608,0.7019608,0.7019608}
\definecolor{colour2}{rgb}{0.8,0.8,0.8}
\pspolygon[linecolor=colour0, linewidth=0.04, fillstyle=solid,fillcolor=colour0](1.2296729,32.603306)(3.094673,28.585804)(3.4546728,27.868305)(2.9346728,26.728304)(2.6096728,26.793304)(2.209673,26.928305)(1.8546729,27.088305)(1.4396728,27.318304)(1.1146729,27.653305)(0.6996729,28.083305)(0.40467286,28.658304)(0.22967285,29.138304)(0.07467285,29.803305)(0.114672855,30.540804)(0.22467285,31.100805)(0.45467284,31.695805)(0.7046729,32.073303)(0.92467284,32.368305)
\pspolygon[linecolor=colour0, linewidth=0.04, fillstyle=solid,fillcolor=colour0](5.792173,32.665806)(6.127173,32.310806)(6.522173,31.680805)(6.817173,31.125805)(6.9471726,30.290804)(6.9071727,29.635805)(6.712173,28.925804)(6.487173,28.335804)(6.099673,27.865805)(5.689673,27.425804)(5.274673,27.145805)(4.869673,26.960804)(4.524673,26.790804)(4.059673,26.725805)(3.489673,27.875805)
\pspolygon[linecolor=colour1, linewidth=0.04, fillstyle=solid,fillcolor=colour1](3.482173,27.848305)(4.0321727,26.698305)(3.5571728,26.683304)(3.1621728,26.703304)(2.9521728,26.708305)
\pspolygon[linecolor=colour1, linewidth=0.04, fillstyle=solid,fillcolor=colour1](1.2046728,32.615803)(1.6046729,32.965805)(2.0046728,33.160805)(2.479673,33.350803)(3.1096728,33.490803)(3.7146728,33.525803)(4.267173,33.405804)(4.737173,33.300804)(5.122173,33.100803)(5.402173,32.953304)(5.7796726,32.690804)(4.4271727,29.875805)(4.462173,29.995804)(4.462173,30.270805)(4.3871727,30.515804)(4.297173,30.670805)(4.127173,30.870804)(3.927173,30.990805)(3.7071729,31.070805)(3.4771729,31.095804)(3.2371728,31.055805)(3.0371728,30.990805)(2.867173,30.870804)(2.6921728,30.700804)(2.6021729,30.550804)(2.5371728,30.365805)(2.4971728,30.100805)(2.5171728,29.910805)(1.7546729,31.508305)
\pspolygon[linecolor=colour2, linewidth=0.02, fillstyle=solid,fillcolor=colour2](9.079673,28.090805)(11.279673,30.290804)(11.879673,29.690804)(9.679673,27.490805)
\pspolygon[linecolor=colour1, linewidth=0.01, fillstyle=solid,fillcolor=colour1](2.6346729,29.575804)(3.479673,27.890804)(4.334673,29.620804)(4.189673,29.475805)(4.064673,29.340805)(3.814673,29.195805)(3.594673,29.125805)(3.2696729,29.155804)(2.979673,29.280804)(2.8396728,29.390804)(2.679673,29.560804)
\pscircle[linecolor=black, linewidth=0.04, dimen=outer](3.479673,30.105804){0.985}
\pscircle[linecolor=black, linewidth=0.04, linestyle=dashed, dash=0.17638889cm 0.10583334cm, dimen=outer](3.504368,30.08989){3.4625}
\psline[linecolor=black, linewidth=0.03](4.2796726,26.190804)(0.77967286,33.490803)
\psline[linecolor=black, linewidth=0.03](2.679673,26.190804)(6.1796727,33.490803)
\pscircle[linecolor=black, linewidth=0.01, fillstyle=solid,fillcolor=black, dimen=outer](3.479673,30.090805){0.07}
\psline[linecolor=black, linewidth=0.02, linestyle=dashed, dash=0.17638889cm 0.10583334cm](3.4746728,30.100805)(3.989673,29.280804)
\psline[linecolor=black, linewidth=0.06, arrowsize=0.05291667cm 2.0,arrowlength=1.4,arrowinset=0.0]{->}(9.079673,26.490805)(9.079673,32.890804)
\psline[linecolor=black, linewidth=0.04, linestyle=dashed, dash=0.17638889cm 0.10583334cm](9.079673,32.490803)(11.879673,29.690804)
\psline[linecolor=black, linewidth=0.04, linestyle=dashed, dash=0.17638889cm 0.10583334cm](9.079673,26.890804)(11.879673,29.690804)
\psline[linecolor=black, linewidth=0.06, arrowsize=0.05291667cm 2.0,arrowlength=1.4,arrowinset=0.0]{->}(9.079673,28.090805)(12.679673,28.090805)
\psline[linecolor=black, linewidth=0.04](9.079673,28.090805)(12.679673,31.690804)
\psline[linecolor=black, linewidth=0.04](9.079673,28.090805)(10.679673,26.490805)
\psline[linecolor=black, linewidth=0.02, linestyle=dashed, dash=0.17638889cm 0.10583334cm](9.079673,29.690804)(11.879673,29.690804)
\psline[linecolor=black, linewidth=0.02, linestyle=dashed, dash=0.17638889cm 0.10583334cm](11.879673,29.690804)(11.879673,28.090805)
\psline[linecolor=black, linewidth=0.02, linestyle=dashed, dash=0.17638889cm 0.10583334cm](3.4846728,30.090805)(1.5946728,27.245804)
\psline[linecolor=black, linewidth=0.04, arrowsize=0.05291667cm 2.0,arrowlength=1.4,arrowinset=0.0]{->}(0.090648465,26.315804)(0.090648465,27.915804)
\psline[linecolor=black, linewidth=0.04, arrowsize=0.05291667cm 2.0,arrowlength=1.4,arrowinset=0.0]{->}(0.090648465,26.315804)(1.4906485,26.315804)
\pscircle[linecolor=black, linewidth=0.01, fillstyle=solid,fillcolor=black, dimen=outer](3.479673,27.855804){0.07}
\psbezier[linecolor=black, linewidth=0.02, arrowsize=0.05291667cm 2.0,arrowlength=1.4,arrowinset=0.0]{->}(1.2396729,28.655804)(1.5096729,28.095804)(1.6346729,28.185804)(1.8196728,28.070804443359375)
\rput[bl](8.7,29.55){$1$}
\rput[bl](3.35,30.25){$0$}
\rput[bl](3.7,29.8){$|y_0|$}
\rput[bl](7.9,26.7){$1-\tau'$}
\rput[bl](7.9,32.3){$1+\tau'$}
\rput[bl](11.7,27.6){$\tau'$}
\rput[bl](9.2,32.6){$y_0$}
\rput[bl](12.4,28.2){$|\vec{y}|$}
\rput[bl](3.74,27.6){$(0,0,-y_0 \tau)$}
\rput[bl](0.5,28.8){$\tau'-|y_0-1|$}
\rput[bl](0.25,27.5){$y_3$}
\rput[bl](1.2,26.5){$y_1,y_2$}

\end{pspicture}
}
\caption{Causal cones in spin dimension one.  On the right, conditions~\eqref{eq:sideconditions} on the coordinates $(y_0,y_1,y_2,y_3)$ representing $\pi_xy\pi_x$ are visualized. On the left, for fixed value of $y_0$ the causal cone in the $(y_1,y_2,y_3)$-hyperplane is shown, using rotational symmetry around the $y_3$-axis.}
 \label{figlightcone2}
\end{figure}
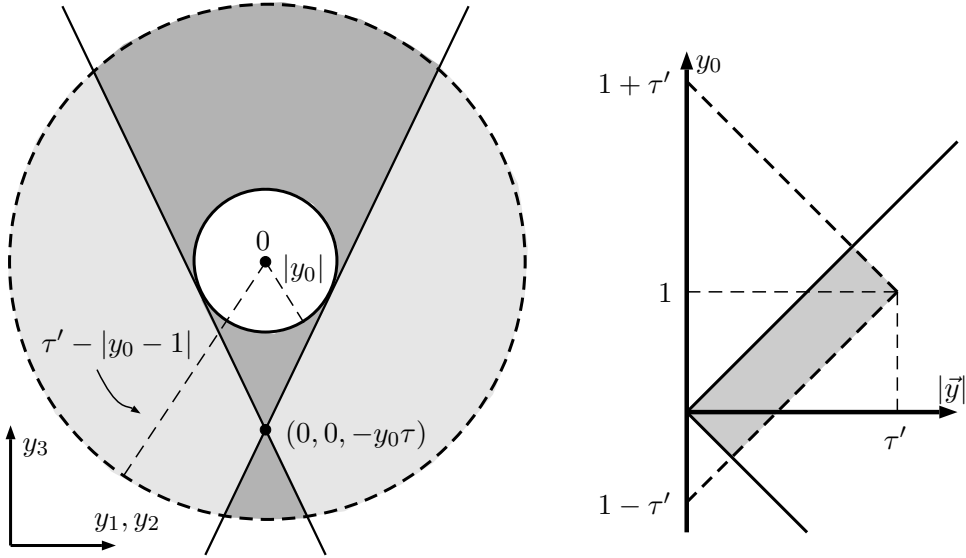%

We finally remark that, using the formula for the eigenvalues~\eqref{eq54}, the causal
Lagrangian~\eqref{Lagrange} can be written as
\beq \label{Lmax}
\L(x,y) = \max \big( 0, \D(x,y) \big) \:,
\eeq
where the function~$\D(x,y)$ is given in analogy to~\eqref{Lform2} by
\beq \label{Ddef}
\D(x,y) = \frac{1}{2} \,\big(\lambda_1 - \lambda_2 \big)^2
= \tr((xy)^2) - \frac{1}{2}\: \tr(xy)^2 \:.
\eeq
This form of the Lagrangian is particularly efficient for numerical calculations.

\section{Numerical Analysis} \label{secnumerics}

\subsection{Overview and Optimization Choices}
In this section we present the numerical optimization scheme.
First, we revisit the causal action principle (see~\eqref{Sdef}) for the weighted counting measure~\eqref{weightcount}.
We seek to solve the optimization problem
\begin{align}\label{optim}
\text{minimize} \quad &\Sact(c_1, \ldots, c_m, x_1, \ldots, x_m) = \sum_{i, j = 1}^m c_i\:c_j\: \L(x_i,x_j) \\
\text{subject to}\quad & c_i \ge 0 \text{ for all } i \in \{1, \ldots, m\}\:, \nonumber\\
                    & \sum_{i=1}^m c_i = 1 \: , \nonumber\\
                    & x_i \in \F \text{ for all } i \in \{1, \ldots, m\}\:, \nonumber
\end{align}
where~$x_i \in \F \subset \C^{f\times f}$ implies that the~$x_i$ are symmetric ($x_i^{*} = x_i$), have unit trace ($\tr x_i = 1$), and at most~$n$ positive and at most~$n$ negative eigenvalues.
Also accounting for the unit-trace requirement we have $\dim_{\R}(\F) = 4 n f - 4 n^2 - 1$~\cite{gaugefix} and thus the parameter space of the optimization problem consists of
\[
D' := m \cdot \dim_{\R}(\F) + m - 1 = m (4 f n - 4 n^2) - 1
\]
real parameters.

In choosing a suitable numerical optimization scheme, we first note that the objective~$\Sact(c_1, \ldots, c_m, x_1, \ldots, x_m)$ is non-convex and not differentiable in the $x_i$.
However, the causal action is continuous in~$c_i$ as well as the entries of~$x_i$ with only finitely many points at which it is not differentiable.
Therefore, we can in principle obtain gradients of the causal action almost everywhere.

Even though local optimization techniques are not guaranteed to converge to a global optimum for this non-convex problem, we prefer them over global schemes due to their computational efficiency in high dimensions.
Specifically, since the number of optimization parameters grows linearly with~$m$ and~$f$ and we are interested in the asymptotics when~$m$ or~$f$ become large, global optimization techniques would quickly become computationally infeasible.
Almost everywhere differentiability of the objective allows us to leverage gradient-information in the optimization.
Recent advances in computational frameworks for gradient-based local optimization such as \texttt{JAX}~\cite{jax2018github} that were mostly developed for machine learning allow for efficient \emph{automatic differentiation}, which we will describe only briefly in the following.

It would be cumbersome to compute the gradient of the objective in~\eqref{optim} analytically.
However, once it is implemented numerically within a framework that allows for \emph{differentiable programming}, the framework traces all computations performed on the input parameters and builds a computation graph of atomic operations that keeps track of how the input parameters were computationally combined to yield the scalar output~$\Sact$.
Atomic operations include arithmetic operations such as addition, subtraction, multiplication, division or powers, as well as elementary functions such as $\log, \exp, \sin, \cos, \ldots$ for which derivatives are known and also implemented within the framework.
If only atomic operations occur in the function implementation, the chain rule can be exploited in an automated iterative fashion (with varying degrees of optimizations) to efficiently compute the gradient of the function at any given input (see, e.g.,~\cite{baydin2018automatic} for more details).
Building on these atomic building blocks, modern libraries can automatically generate gradients even for complex operations such as iterative procedures for eigenvalue computations or matrix factorization routines.
Moreover, they increasingly perform low-level just-in-time compilation with various hardware-specific optimizations to speed up these gradient computations not only for specific central processing unit (CPU) architectures, but also hardware accelerators such as graphics processing units (GPU) or tensor cores~\cite{jax2018github}.

With gradient information about~$\Sact$ readily available via automatic differentiation, we can directly use established optimizers such as BFGS or limited memory BFGS (L-BFGS)~\cite{nocedal2006numerical}.
It has been empirically known and convincingly argued that quasi-Newton methods, in particular BFGS, often converge to local minima for non-smooth, non-convex objective functions~\cite{lewis2009nonsmooth,lewis2013nonsmooth}.
Only recently the seminal proof by Powell of the global convergence of BFGS on smooth convex functions (under mild conditions)~\cite{powell1976some} has been extended to certain non-smooth functions~\cite{guo2018nonsmooth}.
Given the dimensionality of our optimization problem and the structure of the objective, we choose the well-tested BFGS quasi-Newton method for optimization.
The general wisdom has become that L-BFGS, a limited memory version of BFGS, is similarly reliable and greatly accelerates BFGS by maintaining a more memory-efficient representation of the approximated inverse Hessian at each optimization step.
However, the current understanding is that L-BFGS may fail more regularly on non-smooth problems~\cite{asl2021analysis}.
In our empirical evaluation, we combine L-BFGS for the first steps to quickly approach a local minimum and then switch to BFGS.

BFGS only applies to unconstrained optimization problems.
While there is a variant of L-BFGS that can handle box constraints, no existing schemes can directly deal with our unit-sum constraint on~$c$ or the requirement that all~$x_i$ lie in $\F$.
Therefore, we now describe how we implement the causal action numerically step by step as a function of unconstrained optimization parameters.
By construction of this procedure all constraints are satisfied automatically, reducing the degrees of freedom from a larger number of unconstrained optimization parameters back to $D'$.

\subsection{Parametrization of the Optimization Problem}

In the implementation, we optimize over the following \emph{unconstrained} parameters
\[
  \widetilde{c} \in \R^m, \: \mu^{+} \in \R^{m \times n}, \: \mu^{-} \in \R^{m \times n}, \: B^{1} \in \C^{m \times 2n \times 2n}, \: B^{2} \in \C^{m \times 2n \times (f - 2n)}\:,
\]
where we assume that $\Gamma_i := \sum_{l=1}^n (e^{\mu^{+}_{i,l}} - e^{\mu^{-}_{i,l}}) \ge 0$ for each $i \in \{1, \ldots, m\}$ (otherwise we flip the respective superscripts $\pm$).
With these~$D = m(4 f n + 2n + 1) > D'$ (for $n, f, m > 0$) free real parameters we are over-parametrizing the optimization problem in~\eqref{optim}.
The additional degrees of freedom will be removed in the construction of the causal action from these parameters.
In the following, all assignments and conditions for optimization parameters with subscript~$i$ hold for all~$i \in \{1, \ldots, m\}$.
Here and in what follows, the index~$i$ always refers to the numbering of the
points of the weighted counting measure. In order to ensure~$c_i \ge 0$ and~$\sum_{i=1}^m c_i = 1$, we compute the actual weights as
\[
  c_i := \frac{e^{\widetilde{c}_i}}{\displaystyle \sum\nolimits_{i=1}^m e^{\widetilde{c}_i}}\:,
\]
thereby eliminating one degree of freedom.
Next, we compute the at most~$n$ positive and at most~$n$ negative eigenvalues of the operators~$x_i$ from the optimization parameters by
\[
  \nu^{\pm}_{i,j} := \pm \frac{e^{\mu^{\pm}_{i,j}}}{\Gamma_i} \quad \text{for all } i \in \{1, \ldots, m\},\; j \in \{1, \ldots, n\}\:.
\]
This eliminates $m$ degrees of freedom.
Finally, we construct the Hermitian matrices~$x_i$ via uniform transformations
\[
  x_i := U_i \, \diag(\nu_i^+, \nu_i^-, \underset{f - 2n \text{ times}}{\underbrace{0, \ldots, 0}}) \, U_i^{*}\:,
\]
where the unitary matrices~$U_i$ are in turn defined via~$U_i := \exp(-i\, H_i)$ for Hermitian matrices
\[
  H_i :=
    \begin{pmatrix}
      H^{1}_i & B^{2}_i \\[0.2em]
      (B^{2}_i)^{*} & 0
    \end{pmatrix}\,, \quad \text{ where } \quad H^{1}_i := (B^1_i + (B^1_i)^{*}) \odot
    \begin{pmatrix}
      0 & & 1 \\
      & \ddots  & \\
      1 & & 0
    \end{pmatrix}\:.
\]
Here,~$\odot$ denotes element-wise multiplication, i.e.,~$H_i^1$ is
simply the matrix~$B^1_i + (B^1_i)^{*}$ with all diagonal entries set to zero
(these diagonal entries describe irrelevant phase transformations and are therefore disregarded).
Due to this construction, only $2\,n (2\,n - 1) / 2$ entries of~$B_i^1$ remain as actual (complex) degrees of freedom, eliminating a total of $2\,n (2\, n + 1)$ real degrees of freedom.
Note that by construction,~$x_i \in \F$ as the way we chose~$\nu^{\pm}_i$ guarantees that~$\tr x_i = 1$. 
Hence starting from~$D$ real unconstrained optimization parameters, we have eliminated $1 + m + 2\,n (2\, n + 1)$ real degrees of freedom to satisfy all required constraints on~$c$ and~$x_i$, yielding the original number of~$D'$ real degrees of freedom.

Now we can directly compute the Lagrangian for each pair~$x_i, x_j$ via~\eqref{Lagrange} and obtain the causal action by a weighted sum over these terms with weights~$c_i c_j$.
The eigenvalue computation of the products~$x_i x_j$ can be computationally expensive, which is why for~$n=1$ we instead compute the Lagrangian equivalently via the formula (see~\eqref{Lmax} and~\eqref{Ddef})
\[
  \L^{(n=1)}(x_i, x_j) := \frac{1}{2} \max\left\{ 0, 2 \tr((x_i x_j)^2) - \tr(x_i x_j)^2 \right\} \:.
\]

To conclude, setting~$\Lambda := (\widetilde{c}, \mu^{+}, \mu^{-}, B^1, B^2)$ for the collection of all optimization parameters, we can reformulate the constrained optimization problem~\eqref{optim} as the unconstrained optimization problem
\[ \text{minimize } \Sact(\Lambda) \text{ over } \R^D\:, \]
where we independently vary the real and imaginary parts of~$B^1, B^2$.
Once we have implemented the causal action as a function of these~$D$ real parameters, automatic differentiation will provide us with a function that computes the gradient of the causal action at any input.
Hence, provided an initial starting point, we can directly employ local, gradient-based optimization such as (L)-BFGS.

\subsection{Initialization}
Assuming no prior knowledge about the minimizer that we could encode directly into parameter values, we start with random initial parameter values from the following distributions.
We draw the parameters~$\widetilde{c}$ (from which we derive the weights~$c$) independently from a normal distribution with mean~$1$ and small standard deviation~$\sigma_c$ to break potential symmetries.
For the spectra, we allow to select an initial guess~$\mu_0 > 0$ that may be motivated from analytical findings and draw the parameters~$\mu^{+}$ (and~$\mu^{-}$) 
independently from a normal distribution with mean~$\log(\mu_0 + 1/n)$ (and~$\log(\mu_0)$) and standard deviation $\sigma_{\mu}$.
Finally, we independently sample real and imaginary parts of all entries of~$B^1_i, B^2_i$ uniformly from~$(-\pi, \pi)$.
We report details on the parameters~$\sigma_c, \mu_0, \sigma_{\mu}$ as well as other parameters such as tolerances in (L)-BFGS for the different experiments in Appendix~\ref{app:impdetails}.

\section{The Asymptotics for Fixed~\texorpdfstring{$m$}{m} and Large~\texorpdfstring{$f$}{f}} \label{secasyflarge}

\subsection{Analytic Results}
In this section we numerically investigate an asymptotic behavior that is analytically well understood and can serve as a first benchmark for our numerical approach.
To this end, we  fix  the number~$m$ of spacetime points and let the Hilbert space dimension $f$ grow large.
In this limit, the causal action can be bounded from below as follows.
\begin{Prp} \label{lemmalargef}
Assume that
\beq \label{flower}
f \geq mn  \:.
\eeq
Then the minimum the causal action~\eqref{Sdef} is bounded below by
\beq \label{Slower}
\Sact(\rho) \geq \Sact_{\min} := \frac{1}{2m n^3}\:.
\eeq
The lower bound is attained by a weighted counting measure.
\end{Prp}
\Proof
A weighted counting measure that attains the bound can be constructed as follows: Choose all operators $x_i$ in the support of the measure so as to have the non-trivial eigenvalues (counting multiplicities) 
\[ \nu_1 = \cdots \nu_n = \frac{1}{n} \qquad \text{and} \qquad \nu_{n+1} \cdots \nu_{2n} = 0 \:. \] 
Thus the image of each operator is only $n$-dimensional, and under the assumption~\eqref{flower} the operators~$x_1,\ldots,x_m$ can then be chosen to have pairwise orthogonal images.
Then~$\L(x_i, x_j)$ vanishes unless~$i=j$. 
The eigenvalues of the operator product~$x^2$ are~$1/n^2$ and zero. Therefore,
\[ \L(x_i, x_i) = \frac{1}{4n} \:2 \:\sum_{i=1}^n \sum_{j=n+1}^{2n} \nu_i^4
= \frac{1}{2n} \:\frac{n^2}{n^4} = \frac{1}{2n^3} \:. \]
We next choose all the weights equal to~$c_i = 1/m$. We thus obtain
\[ \Sact(\rho) = \frac{1}{m^2} \sum_{i=1}^m \L(x_i, x_i) = \frac{1}{2n^3\, m}\:, \]
showing that the value~$\Sact_{\min}$ in~\eqref{Slower} is attained.

In order to show that this value of the causal
action is minimal, we slightly improve the estimate in~\cite[Proposition~4.3]{discrete}: We denote the non-trivial eigenvalues of~$x$ by~$\nu_1, \ldots, \nu_{2n}$ and order them such that
\[ \nu_1, \ldots \nu_n \geq 0 \qquad \text{and} \qquad \nu_{n+1}, \ldots, \nu_{2n} \leq 0 \:. \]
Then, omitting the summands in~\eqref{Lagrange} for which both~$i,j \leq n$ or both~$i,j > n$, we obtain
\begin{align*}
\L(x,x) &\geq \frac{1}{4n} \:2 \:\sum_{i=1}^n \sum_{j=n+1}^{2n} \big( \nu_i^2 - \nu_j^2 \big)^2
= \frac{1}{2n} \:\sum_{i=1}^n \sum_{j=n+1}^{2n} \big( \nu_i - \nu_j \big)^2 \big( \nu_i + \nu_j \big)^2 \:.
\end{align*}
In this formula, the eigenvalues~$\nu_i$ and~$\nu_j$ have opposite sign. This gives the estimate
\beq \label{Llower}
\L(x,x) \geq \frac{1}{2n} \:\sum_{i=1}^n \sum_{j=n+1}^{2n} \big( \nu_i + \nu_j \big)^4 \:.
\eeq
On the other hand, the trace of~$x$ can be estimated with the help of the H\"older inequality by
\[ n = n\: \tr x = \sum_{i=1}^n \sum_{j=n+1}^{2n} \big( \nu_i + \nu_j \big) 
\leq \bigg( \sum_{i=1}^n \sum_{j=n+1}^{2n} \big( \nu_i + \nu_j \big)^4 \bigg)^\frac{1}{4} \:
\big( n^2 \big)^\frac{3}{4}\:. \]
Using this inequality in~\eqref{Llower}, we obtain the lower bound
\[ 
\L(x,x) \geq \frac{1}{2n^3} \:. \]
This makes it possible to estimate the causal action by
\[ \Sact(\rho) \geq \sum_{i=1}^m c_i^2\: \L(x_i, x_i) \geq \frac{1}{2n^3}
\sum_{i=1}^m c_i^2 \:. \]
Using the Cauchy-Schwarz inequality
\[ 1 = \sum_{i=1}^m c_i \leq \Big( \sum_{i=1}^m c_i^2 \Big)^\frac{1}{2}\: \sqrt{m} \:, \]
we obtain the desired lower bound in~\eqref{Slower}.
\QED
Specializing this result to the cases~$n=1$ and~$n=2$, we find that for
\begin{align*} 
\text{large $f$ at $n=1$:}&\qquad \min \Sact \overset{f \to \infty}{\longrightarrow} \frac{1}{2 m} \\
\text{large $f$ at $n=2$:}&\qquad \min \Sact \overset{f \to \infty}{\longrightarrow} \frac{1}{16 m} \:.
\end{align*}

Having established this optimal lower bound for the causal action in the limit~$f\to\infty$, we
now state another lower bound for a specific class of measures in spin dimension $n=1$.

\begin{Lemma}\label{lemma:n1Sbound}
Let $\rho=\sum_{i=1}^m \frac1m\delta_{x_i}$ be a normalized equal-weighted counting measure on $\F$ in spin dimension $n=1$ for which all operators~$x_i$ have rank one. Then
\begin{align}\label{eq:S_n1_minfty}
\Sact (\rho)\geq\frac1{f(f+1)}.
\end{align}
\end{Lemma}
\Proof
Since the operators~$x_i$ have rank one and~$\tr x_i=1$,
all the~$x_i$ are projection operators to one-dimensional subspaces in~$\H$, i.e., in
bra/ket notation,
\[ x_i=\ketbra{v_i}{v_i} \qquad \text{with unit vectors~$v_i \in \H$}\:. \]
As a consequence, the product of two operators $x_ix_j$ has rank at most one, and its non-trivial eigenvalue
is given by~$\left|\braket{v_i}{v_j}\right|^2$. Hence
\[ \L(x_i,x_j)=\frac{\left|\braket{v_i}{v_j}\right|^4}2 \:. \]
The (second) Welch bound~\cite{welch} gives the estimate
\[ \sum_{i,j=1}^m \left|\braket{v_i}{v_j}\right|^4\geq \frac{m^2}{\displaystyle \binom{f+1}{2}}=\frac{2m^2}{f(f+1)} \:. \]
We thus obtain the lower bound
\[ \Sact(\rho)=\sum_{i,j} c_i \,c_j\,\L(x_i,x_j)=\frac1{2m^2}\sum_{i,j}\L(x_i,x_j)\geq \frac1{f(f+1)} \:. \]
This concludes the proof.
\QED

We point out that the last estimate applies only if all the operators have rank one.
At least for very large~$m$, this is not the case for minimizing measures, as the following argument shows.
We choose~$k$ such that~$2k \leq f \leq 2k+1$. Then we can construct $k$ copies of a discrete Dirac sphere
on~$k$ pairwise orthogonal, two-dimensional subspaces of $\H$. Then, in the limit~$m\to\infty$, the 
causal action tends to
\[ \Sact\to\frac1{k^2}\frac{\sqrt3}{4\pi} \:. \]
If~$f$ is even, we conclude that~$\Sact \rightarrow \sqrt3/(\pi f^2)$, which is smaller than the
lower bound in~\eqref{eq:S_n1_minfty}. Likewise, if~$f$ is odd, we obtain~$\Sact \rightarrow \sqrt3/(\pi (f-1)^2)$,
which is again smaller than the lower bound in~\eqref{eq:S_n1_minfty} if~$f \geq 7$.

This consideration shows that the assumptions of Lemma~\ref{lemma:n1Sbound}
are not satisfied asymptotically as~$m \rightarrow \infty$. 
Nevertheless, the lower bound~\eqref{eq:S_n1_minfty}
will be helpful for interpreting our numerical results in the case~$m>f$, as we will
see in the next section.

\subsection{Numerical Findings}
\begin{figure}
  \centering
  \includegraphics[height=4.5cm]{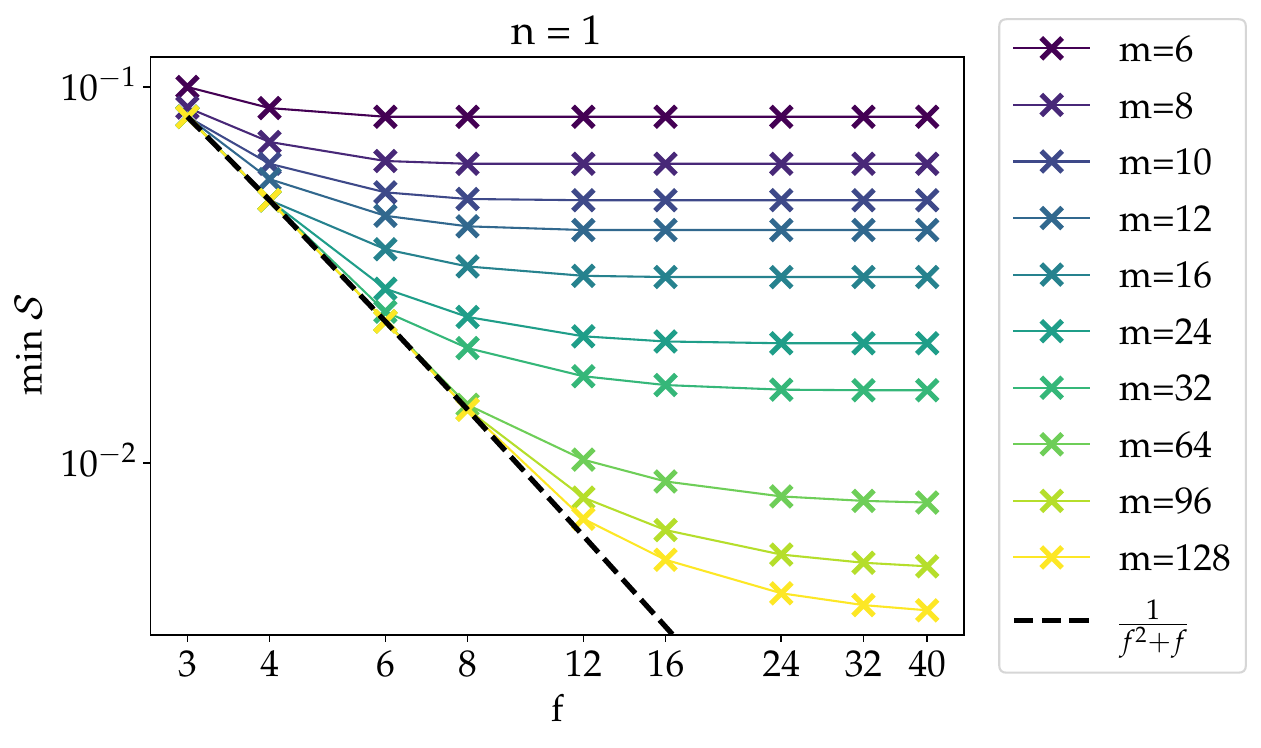}
  \hfill
  \includegraphics[height=4.5cm]{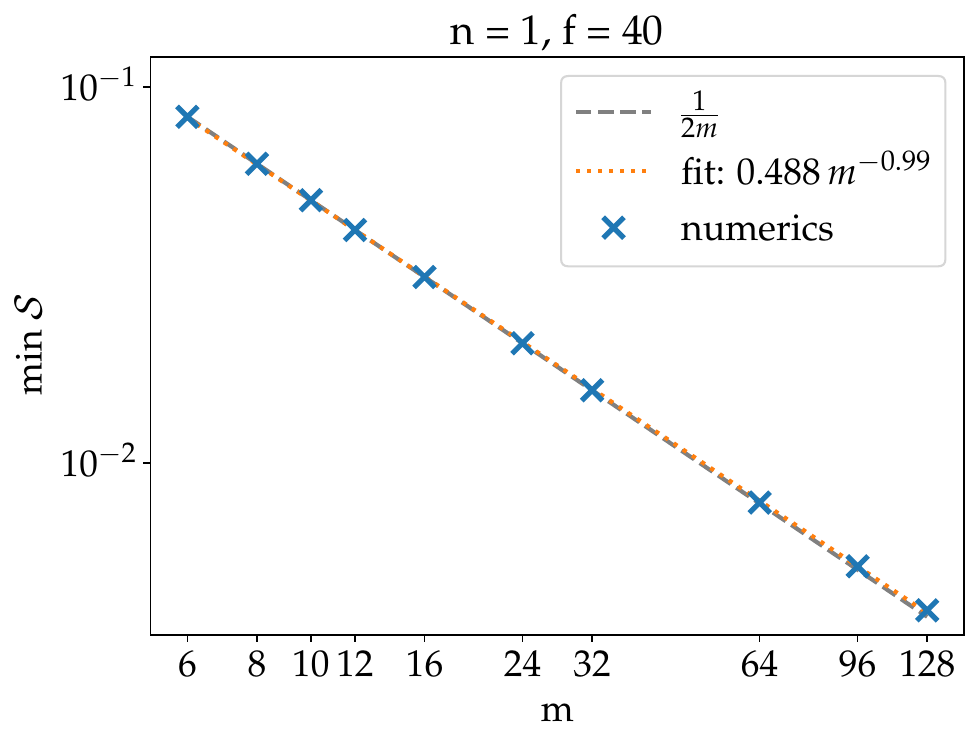}\\
  {\color{gray}\hrule}
  \vspace{2mm}
  \includegraphics[height=4.5cm]{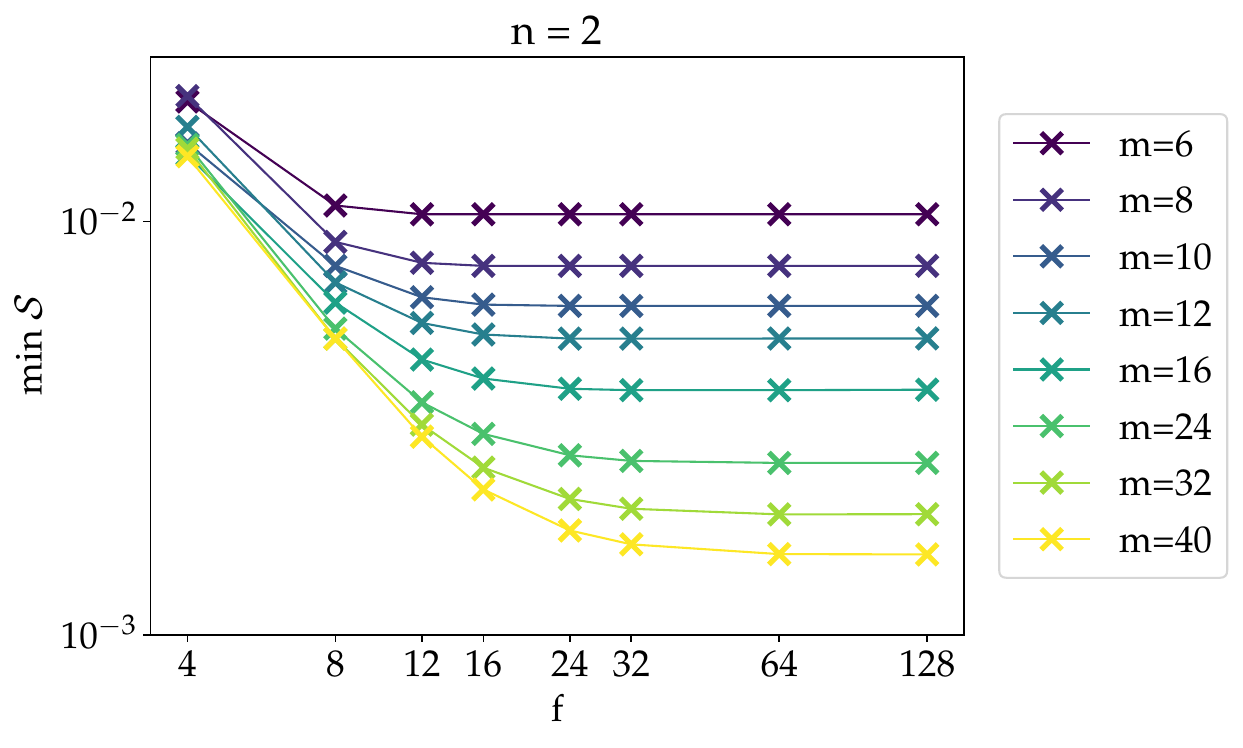}
  \hfill
  \includegraphics[height=4.5cm]{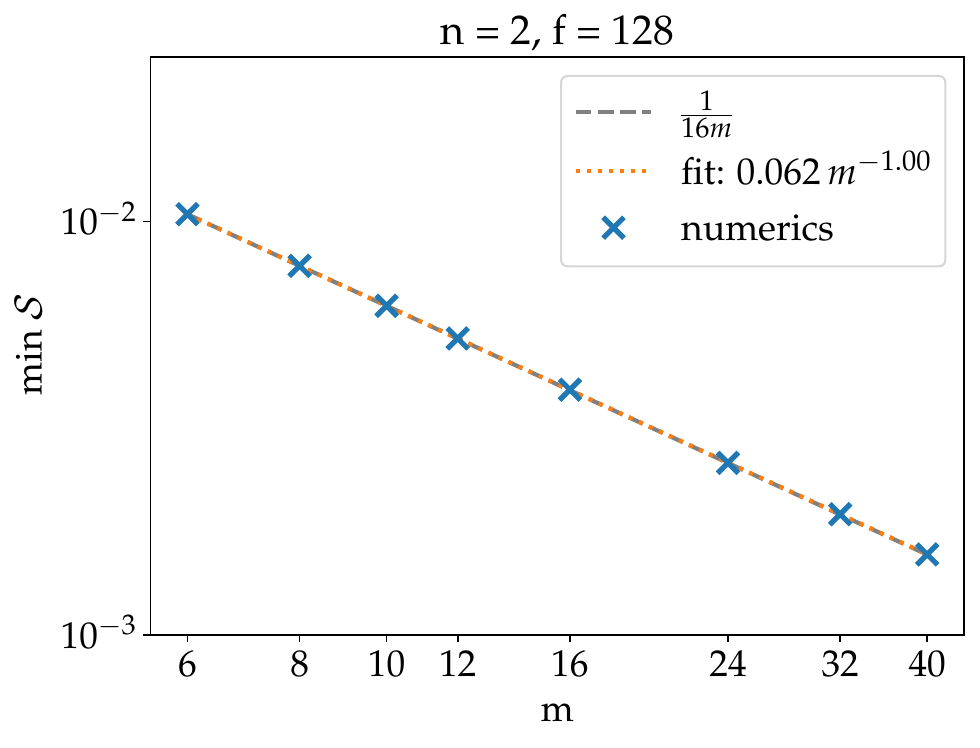}
  \caption{In the left column we show the numerically found minimal causal action for increasing values of~$f$ and~$m$ at~$n=1$ (top) and~$n=2$ (bottom).
  The right column then displays the scaling behavior of the minimal causal action at the largest value of~$f$ for each~$m$.
  A simple linear least squares fit (in log-log space) confirms the analytic asymptotic behavior of~$\Sact$ as~$f$ becomes large for both~$n=1$ and~$n=2$.}
  \label{fig:self_contribution}
\end{figure}

Figure~\ref{fig:self_contribution} shows that our numerical results reproduce the analytic asymptotics to great accuracy.
We take this as first evidence that the numerical optimization scheme appears to find these simple minimizing configurations for various combinations of values of~$m$ and~$f$, even for $D = 41\,160$-dimensional optimization problems.
All these results were found from a single random starting point.

Whereas this section focuses on  the large $f$ limit, the upper row in Figure~\ref{fig:self_contribution} for spin dimension $n=1$ also shows data for $m>f$.
Here we see that the numerically found configurations only exceed the $S\propto1/(2m)$ scaling of the causal action by very small values, despite the violation of the assumption of Proposition~\ref{lemmalargef}.
This is not surprising in view of Lemma~\ref{lemma:n1Sbound}, 
which shows that the value~$\Sact(\rho)=1/(f^2+f)$ can be achieved if there exists a choice of $m$ unit vectors in $\mathbb{C}^f$ attaining equality in the relevant  Welch bound.
In fact,  the upper left plot in Figure~\ref{fig:self_contribution} shows that for $m>f$, the numerically found configurations follow this asymptotic behavior. We also find for 
these numerical configurations that all spacetime point operators are (within the numerical accuracy)
projection operators of rank one.
Nevertheless, as discussed after Lemma~\ref{lemma:n1Sbound}, the causal action can be lowered below $\Sact=1/(f^2+f)$ for sufficiently large $m$. We do not see this effect in our numerics, however, probably because the
considered values of~$m$ are not large enough.
%

\section{Numerical Study of Discrete Dirac Spheres} \label{secdiracnum}
In this section we study the asymptotic behavior of the causal action when the Hilbert space dimension is kept minimal at~$f=2n$ but the number~$m$ of spacetime points grows. In both cases~$n=1$ and~$n=2$,
our results follow the asymptotic behavior expected from the analytical results on
discrete Dirac spheres from Sections~\ref{secdirac2d} and~\ref{secdirac4d}.
The numerically found causal actions come close but do not get below the values
calculated there.
\begin{figure}
  \centering
  \includegraphics[height=5cm]{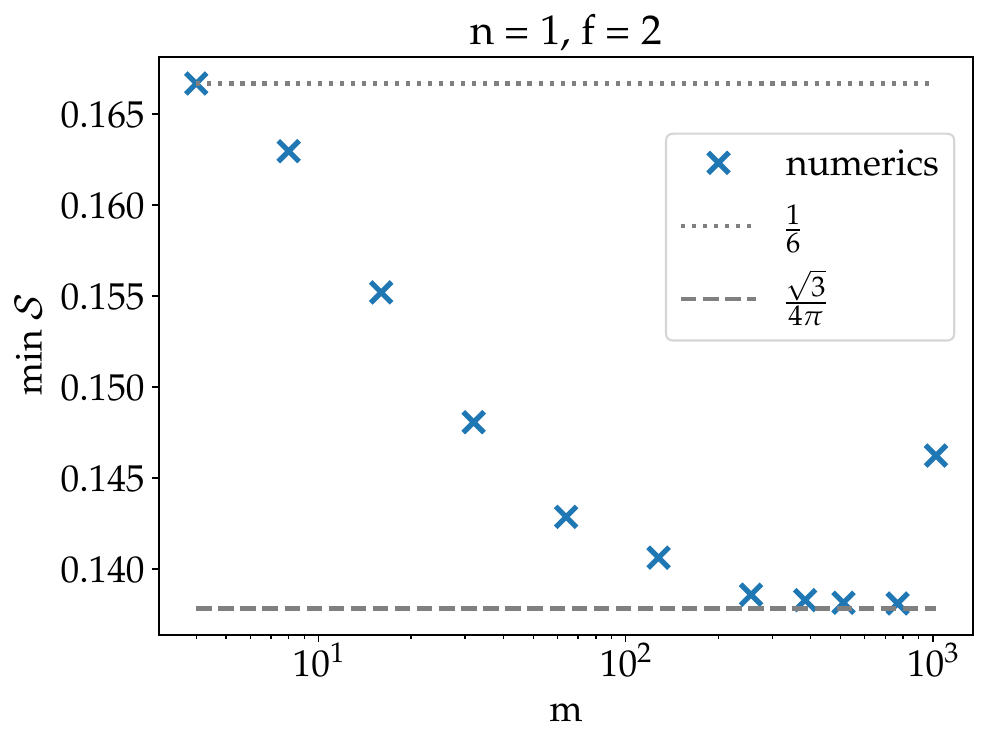}
  \hfill
  \includegraphics[height=5cm]{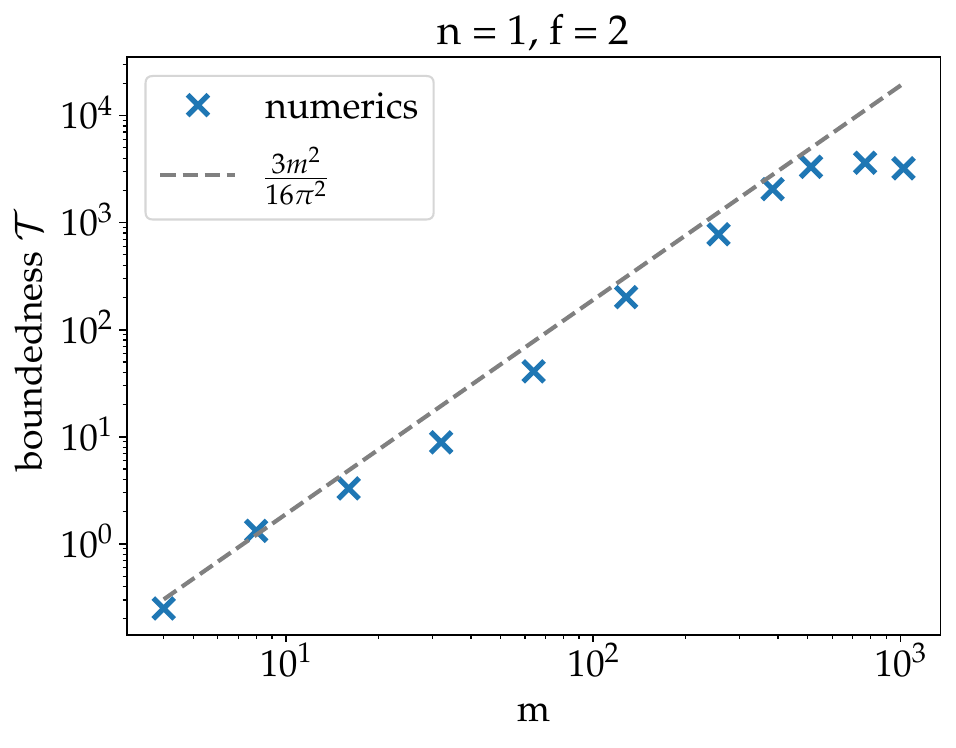}\\%
  {\color{gray}\hrule}
  \vspace{2mm}
  \includegraphics[height=5cm]{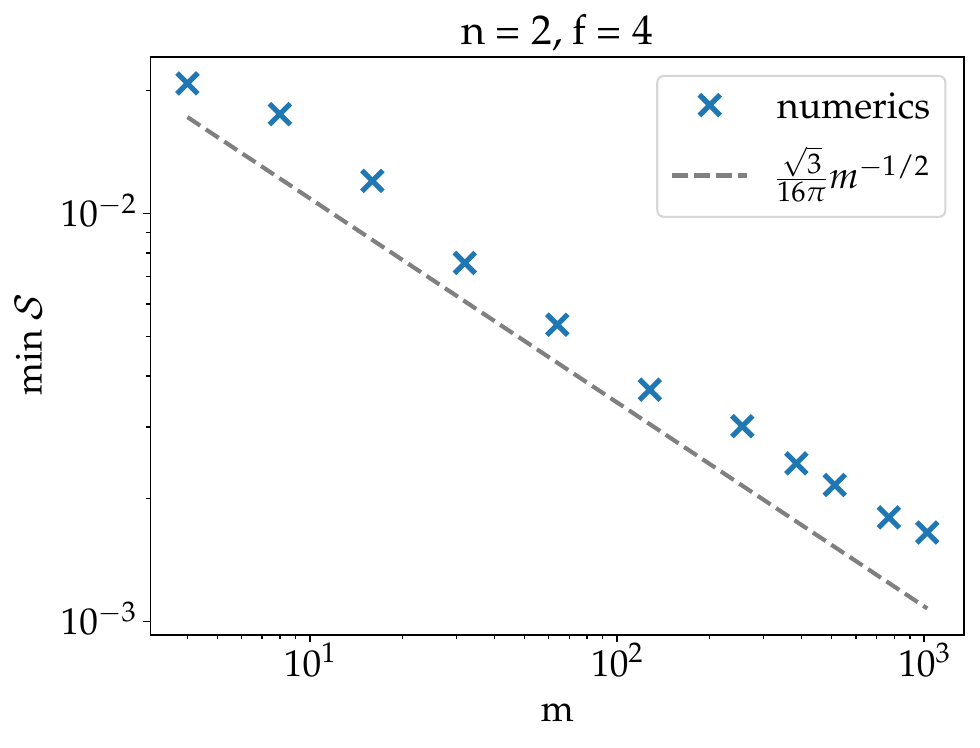}
  \hfill
  \includegraphics[height=5cm]{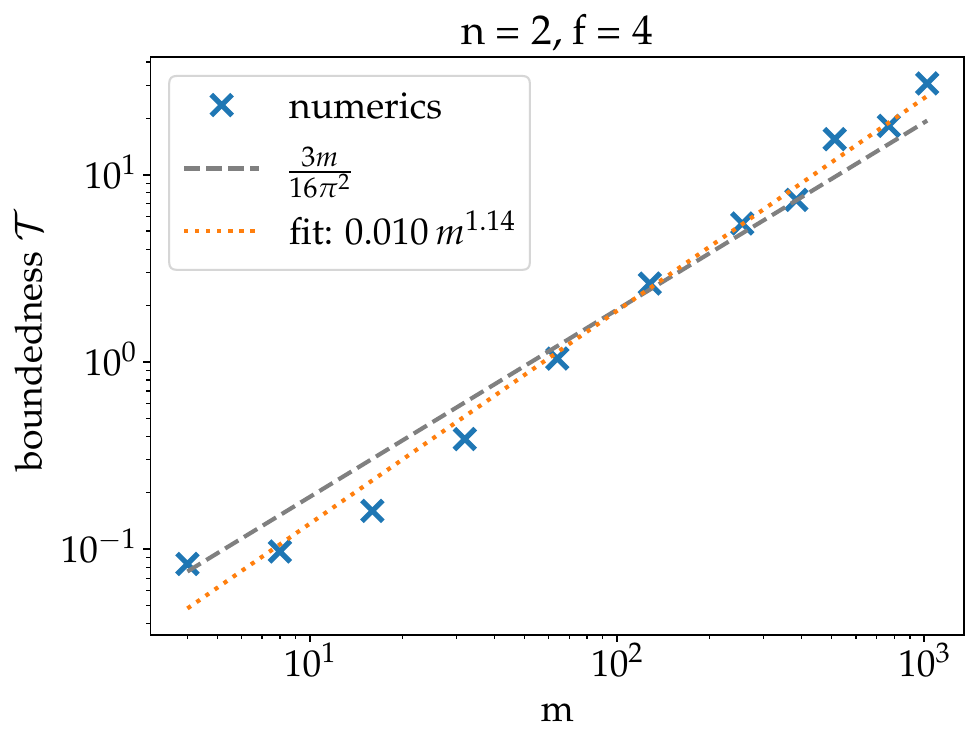}
  \caption{The minimal causal action (left) and value of the boundedness functional (right) found numerically for $n=1, f=2$ (top row) and $n=2, f=4$ (bottom row) as $m$ increases.
}
  \label{fig:dirac_sphere}
\end{figure}

For the numerical results of  this section, each individual optimization was repeated for ten different random starting points, and results are reported only for the run with the smallest final value of~$\Sact$.

\subsection{Spin Dimension One}
In the case~$f=2$, Proposition~\ref{propositionn1} shows that for the two-dimensional discrete Dirac sphere the causal action approaches~$\Sact \to \frac{\sqrt{3}}{4 \pi}$ as~$m \to \infty$ for~$\tau \propto \sqrt{m}$, and~\eqref{eq:iso_measure_bound} in Proposition~\ref{propositionn1_fixedtau} shows that for an isotropic distribution of spacetime points at fixed~$\tau = 1$ the causal action approaches $\Sact \to \frac{1}{6}$ as~$m \to \infty$.
The top row in Figure~\ref{fig:dirac_sphere} shows the results of our numerical optimization.
As~$m$ increases, the minimal causal action indeed drops from the upper bound at~$\frac{1}{6}$ for small~$m$ to the analytical minimum of Proposition~\ref{propositionn1} for~$m \approx 1000$.
We note that for large~$m$ the dimensionality of the optimization problem is of the order of~$10^4$,
and we stopped all optimization runs after 3 days, which is why runs for~$m=1024$ have not fully converged.

\begin{figure}
    \begin{center}
        \input{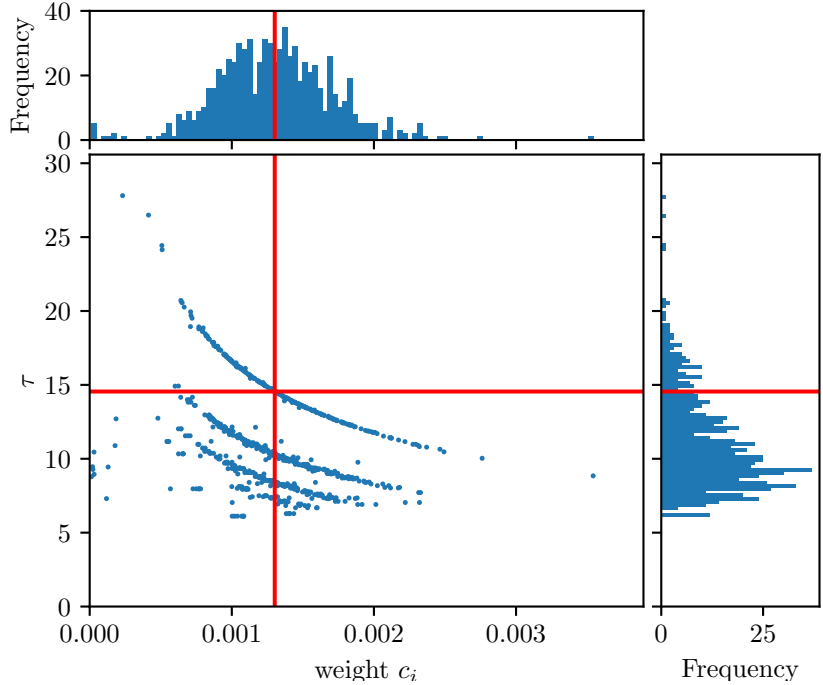}
    \end{center}
    \caption{Clustering in the numerical configuration for $n=1,f=2,m=768$. Red lines indicate expectations from discrete Dirac sphere.}
\label{figclustering}
\end{figure}

The right plot in the top row of Figure~\ref{fig:dirac_sphere} shows that the asymptotic growth rate of the boundedness functional at first is quadratic for large~$m$ and closely matches the analytic results from Proposition~\ref{propositionn1T}, i.e., $\T =  \frac{3}{16 \pi^2} m^2 +\O(m)$. However, for the largest values of $m$, 
it deviates from the expected scaling.
This can be explained by the fact, that in the numerical configurations the values of $\tau_i$ have a certain spread where most $\tau_i$ may even lie significantly below the value expected from the discrete Dirac sphere. This is
shown in Figure~\ref{figclustering}.

This figure also reveals another surprising feature of the numerically found configurations.
When plotting~$\tau_i$
versus the weight~$c_i$ of the spacetime points of a numerical minimizer,
one sees that the points tend to accumulate on several hyperbolas with
\[ \tau_i^2\,c_i = \text{const}. \]
The first, upper-most of these hyperbolas goes through the point~$(\tau, c)$ corresponding to the 
parameter values of the discrete Dirac sphere (see the red cross hair in the figure) in Section~\ref{secdirac2d}.
In fact, the spacetime points belonging to this hyperbola are (essentially) all pairwise spacelike separated from each other. Hence, they can be understood as a discrete Dirac sphere configuration which is causally trivial
(as defined and analyzed in Section~\ref{sectivcausal}), in which the $\tau_i$ vary and the weights $c_i$ are adjusted according to~\eqref{civals}. 
The spacetime points on the other hyperbolas, however, are not causally trivial.
Instead, on the first hyperbola below the causally trivial hyperbola, every spacetime point is timelike separated to itself and to one other spacetime point.
Similarly, on the second hyperbola every spacetime point has two partners with timelike separation.
Generally speaking, the spacetime points form groups with timelike separation within each group but spacelike separation between points in different groups.
The points within each group are in a quite simple configuration: they have almost the same value of~$\tau$,
have almost the same weight, and the corresponding vectors~$\vec{x}$ in the representation~\eqref{eq:R3DiracMap} are very close to each other.
Therefore, each group can be viewed as a ``cluster'' of nearby points on~$\F$ with equal weight factors.
If one replaces each cluster by a single point (with the weight taken as the sum of the weights of the
cluster points), then the resulting spacetime point lies again on the uppermost, causally trivial hyperbola
in Figure~\ref{figclustering}.
In this way, we obtain a numerical configuration which coincides with a discrete Dirac sphere configuration,
except that the values of~$\tau$ are smeared out. Keeping in mind that the numerical values of the
causal action always lie slightly above the value for the discrete Dirac sphere, our findings
support the conjecture that the absolute minimizers for large~$m$ should indeed be discrete Dirac sphere
configurations. We expect that the numerical configurations with more spread out values of~$\tau$
correspond to local minima.

\subsection{Spin Dimension Two}
For~$f=4$, Proposition~\ref{propositionn2} shows that for the four-dimensional Dirac sphere~$\Sact = \frac{\sqrt{3}}{16 \pi \sqrt{m}} + \O(m^{-\frac{3}{2}})$ for~$\tau \propto m^{\frac{1}{4}}$.
The bottom row in Figure~\ref{fig:dirac_sphere} shows the results of our numerical optimization.
While the found configurations do not reproduce the analytical values for the minimal causal action precisely, they are reasonably close and---more importantly---reproduce the correct scaling behavior.
We remark once again that these are non-convex, non-smooth optimization problems in high dimensions ($D=37\,888$ for~$m=1024$) tackled with local quasi-Newton methods.
The analytic asymptotic scaling behavior of the boundedness functional for large~$m$ from Proposition~\ref{prpT4} in leading order becomes $\T = \frac{3}{16 \pi^2}\, m$ when substituting $\tau$ from~\eqref{tauval4}.
This value is reproduced by our numerics with good accuracy.

To summarize, the numerics presented in this section yield strong evidence for the conjecture that
minimizers of the causal action principle for large~$m$ are close to
the two- and four-dimensional discrete Dirac spheres, respectively.

\section{Numerical Study beyond Discrete Dirac Spheres} 
\label{secf3}
At present, it is an open problem to understand the asymptotic behavior of the causal action
for a large number~$m$ of spacetime points for fixed Hilbert space dimension~$f$.
In the previous sections, we found that 
our numerical results support our analytic understanding and confirm conjectures on the behavior of minimizers in the minimal Hilbert space dimension~$f=2n$ (again for large~$m$).
However, beyond this case of minimal Hilbert space dimension, no analytic candidates for asymptotic minimizers are known. Therefore, it is an important task to use our numerical methods in order
to explore and understand these cases better.

In the case of spin dimension $n=2$, already in the minimal case $f=4$ we showed that the causal action
tends to zero as for large~$m$ at least at the rate~$\Sact\propto \frac1{\sqrt m}$.
Since the infimum of the causal action is monotone decreasing in~$f$,
the causal action also tends to zero for if $f>4$, but the rate of decay in~$m$ could be faster.
In the case of spin dimension~$n=1$, the situation is even less clear, because it is not known
if there are measures that achieve arbitrary low values of the causal action as~$m\to\infty$, or if,
conversely, for each fixed~$f$ there exists a strictly positive lower bound of the causal action.

In the case~$f=2$, the numerical results seem to support the conjecture that all minimizers for
large~$m$ are close to the two-dimensional Dirac sphere, so that~\eqref{Sasy} provides a
universal lower bound for the causal action.
However, already in the case~$n=1, f=3$ no analytic candidate for a minimizer is known 
asymptotically for large~$m$. The basic difficulty can be understood as follows. 
In the case~$n=1, f=2$,
the set~$\F$ can be parametrized by linear combinations of the identity matrix and the three Pauli matrices
(see~\eqref{eq:R3DiracMap}). Here a natural candidate for minimizers is the discrete Dirac sphere
(see Section~\ref{secdirac2d}). The fact that the anti-commutation relations of the Pauli matrices
play a special role in the computation of the causal Lagrangian motivates higher-dimensional
discrete Dirac spheres, like the four-dimensional discrete Dirac sphere introduced in Section~\ref{secdirac4d}.
Such discrete Dirac sphere configurations exist only for specific choices of~$f$ and~$n$.
This follows from the representation theory of Clifford algebras
(for details see for example~\cite{lawson+michelsohn} or~\cite{snygg}). More specifically, $f$
must be the dimension of an irreducible Euclidean Clifford representation, giving rise to the
\[ \text{$2k$-dimensional discrete Dirac sphere for} \qquad \text{$n=2^{k-1}$ and~$f=2^k$} \]
with~$k \in \N$. Since there is no irreducible Clifford representation on~$\C^3$,
in the case~$f=3$ there is no discrete Dirac sphere.

In view of this lack of analytic candidates,
the numerical investigation of the case~$n=1$ and~$f=3$ can serve the important
task of building intuition and gaining a better understanding of the structure of minimizers.
To this end, it is helpful to represent the numerically found discrete spacetime in a way which gives
insight into its geometric and causal structure. This is accomplished by the 
{\emph{projected spacetime plot}}, which visualizes the spacetime as seen from a chosen reference point, 
based on the geometric cone structures uncovered in Section~\ref{seccausalgen}.

\begin{figure}
    \begin{center}
        \input{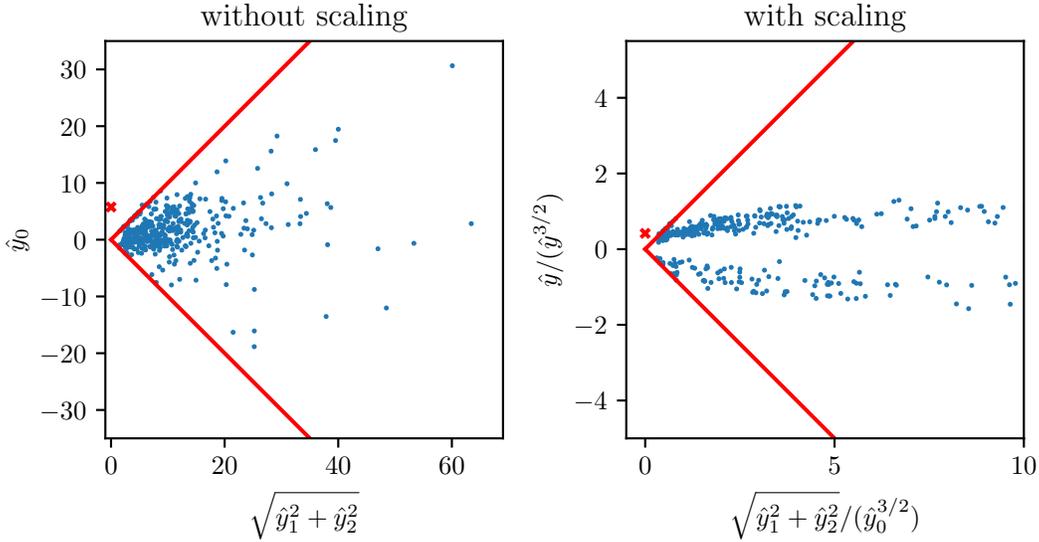}
    \end{center}
    \caption{Projected spacetime plot, without and with the scaling~\eqref{eq:rescale}, for a numerical configuration with $n=1,f=3,m=512$. The red lines indicate the causal cone of the reference point, whose position inside the cone is marked by a cross.}
\label{fig:feq3}
\end{figure}

The projected spacetime plot is obtained as follows.
We let~$(c_i, x_i)_{i=1,\ldots, m}$ with~$c_i \geq 0$ and~$x_i \in \F$ be a numerical minimizer.
We choose one point~$(c_j, x_j)$ with~$j \in \{1, \ldots, m\}$ as our reference point.
For consistency with the notation in Section~\ref{seccausalgen}, we denote this reference point by~$x=x_j$.
Then the image of the spacetime point operator~$x(\H)$ is a two-dimensional subspace of~$\H$
(this subspace is the so-called spin space of the spacetime point; for details see
for example~\cite[Section~1.1]{cfs} or~\cite[Section~5.6]{intro}).
For convenience, we work in an eigenvector basis of~$x(\H)$, where our reference point
has the matrix representation~\eqref{xbasisrep}.
Next, we let~$y=x_i$ be any spacetime point. Using that the Lagrangian~$\L(x,y)$ can be
computed from the operator product~$x \,\pi_x y \pi_x$,
we consider, instead of~$y$, the projection~$\pi_x y \pi_x$ of~$y$
to~$x(\H)$ and represent it according to~\eqref{eq:matrix_PyP}
in terms of real coordinates~$(y_0, \ldots, y_3)$.
We want to plot these coordinates for all spacetime points~$x_1, \ldots, x_m$ in such a way
that the causal structures as well as the distances between these points become visible.
In view of~\eqref{relcausal}, plotting the coordinates
\[ \hat{y}_0 := y_3 + y_0 \tau \qquad \text{and} \qquad
\hat{y}_{1\!/\!2} := \sqrt{\tau^2-1}\: y_{1\!/\!2} \:, \]
the causal relations between the reference point~$x$ and the spacetime point~$y$
coincide with that of three-dimensional Minkowski space~$\R^{1,2}$ with the metric
\[ \big\la (\hat{y}_0, \hat{y}_1, \hat{y}_2) , (\hat{y}_0, \hat{y}_1, \hat{y}_2)  \big\ra_{\R^{1,2}}
:= \hat{y}_0^2 - \hat{y}_1^2 - \hat{y}_2^2 \:. \]
In order to simplify the plot, we make use of the fact that the causal Lagrangian~$\L(x,y)$ is
rotationally symmetric in the~$\hat{y}_1-\hat{y}_2$-plane. This is also reflected by
our numerical minimizers which do seem to distinguish a direction in this plane.
Therefore, it is preferable to plot~$\hat{y}_3$ versus~$\sqrt{\hat{y}_1^2+\hat{y}_2^2}$,
as shown on the left of Figure~\ref{fig:feq3}.

In order to get the correct intuition about distances, one should keep in mind that the projections~$\pi_x$
make the parameters~$(y_0, \ldots, y_3)$ smaller for points which are to be thought of as being further apart from~$x$. More specifically, one can argue as follows. If the parameters~$(y_0, \ldots, y_3)$ are large
and the points~$x$ and~$y$ are timelike separated, then the Lagrangian~$\L(x,y)$ is typically large.
In concrete examples, the Lagrangian decays if the spacetime points are far apart.
Therefore, points for large~$\L(x,y)$ should correspond to points which are close together.

This picture becomes more precise with the following scaling argument. Having two spatial dimensions
in mind, Dirac wave functions have the length dimension minus one
(because $\int_{\R^2} |\psi| d^2 x$ is a dimensionless probability). Since the local correlation operators
involve two wave functions, they have the length dimension minus two. Therefore, in order to obtain
Minkowski vectors with the dimension of length, one should plot the vectors
\begin{align}\label{eq:rescale}
\frac{1}{|\hat{y}_0|^\frac{3}{2}}\: \big( \hat{y}_0,\: \hat{y}_1,\:  \hat{y}_2 \big) \:. \end{align}
This gives plots as shown on the right of Figure~\ref{fig:feq3}.

In this way, the causal fermion system can be visualized by a discrete subset of Minkowski space.
The causal structure of Minkowski space coincides with the causal relations between
the reference point~$x$ and all the other spacetime points. Moreover, the size of the Minkowski vector
gives an indication of the ``distance'' between~$x$ and~$y$.

The numerical data obtained for~$n=1, f=3$ for spacetime point numbers up to~$m=768$, 
obtained within the scope of this work, 
do not yield conclusive results for a potential lower bound of the causal action.
To this end, it appears necessary to increase~$m$ by at least one order of magnitude. This, in turn,
makes it necessary to improve the efficiency of our numerical approach.

\section{Conclusion and Outlook} \label{secconclusion}
The main conclusion of the present work is to demonstrate that
differentiable programming methods give a major advance in the numerical study
of the causal action principle for causal fermion systems.
We gave a detailed analysis in the cases~$n=1, f=2$, $n=1, f=3$
and~$n=2, f=4$.
In the cases~$n=1, f=2$ and~$n=2, f=4$, our results support the conjecture that,
asymptotically for large~$m$, the absolute minimizers should be close to discrete Dirac sphere configurations.

Our analysis gave new insight into the structure of minimizing measures. In particular, for spin dimension~$n=1$, the causal relations could be formulated geometrically in terms of causal cones. Moreover, as a helpful tool to build intuition for the structure of numerically found minimizers,
we introduced projected spacetime plots which visualize the causal and geometric structure of the resulting discrete spacetimes as seen from an arbitrary reference point.

We see several directions for future research. First, we plan to
complement the numerical toolbox by linear programming methods in convex domains
(in this case, one approximates~$\F$ by a sufficiently large discrete set and minimizes the
causal action under the variation of the weights in the convex set~$\{ 0 \leq c_i \text{ and } \sum_i c_i=1\}$).
\Felix{Nehme diesen Satz eventuell heraus?}%
Another open problem concerns the asymptotics of the minimal causal action for fixed~$n$ and~$f$
and large~$m$ (except for the cases~$n=1,f=2$ and~$n=2,f=4$, which are already well-understood).
Finally, the major goal of future research will be to analyze the case of larger~$f$ numerically.
This is a challenging problem, because the number of degrees of freedom grows linearly
in~$f$. Therefore, an efficient numerical study requires more sophisticated methods which are more
adapted to the structure of a causal action principle.
We plan to develop these methods in the near future.

\Thanks{{{\em{Acknowledgments:}} 
R.H.J. gratefully acknowledges support by the Wenner-Gren Foundations. }

\appendix
\section{Details of the Implementation} \label{app:impdetails}

All our code is implemented in Python 3.9~\cite{vanrossum2009python} making use of \texttt{JAX}~\cite{jax2018github}, Numpy~\cite{harris2020numpy}, Scipy~\cite{harris2020numpy}, Matplotlib~\cite{hunter2007matplotlib} for visualizations, as well as abseil-py\footnote{\url{https://github.com/abseil/abseil-py}} and tqdm\footnote{\url{https://github.com/tqdm/tqdm}}.

Throughout all our experiments we fix the parameters~$\sigma_c = \sigma_{\mu} = 0.01$ and choose $\mu_0$ as a function of~$m$ following Proposition~\ref{propositionn1}
\begin{equation*}
\mu_0^{(n=1)}(m) = \frac{5}{4} \left(3^{\frac{1}{4}} \sqrt{\frac{m}{2 \pi}} - 1\right)
\end{equation*}
for all experiments with $n=1$ and following Proposition~\ref{propositionn2}
\begin{equation*}
\mu_0^{(n=2)}(m) = \frac{1}{4} \left(\frac{(3 m)^{\frac{1}{4}}}{\sqrt{\pi}} - 1\right)
\end{equation*}
for $n=2$.

We use the Scipy implementations of (L)-BFGS.
We start the optimization with L-BFGS and parameters \texttt{ftol}${}=10^{-7}$, \texttt{gtol}${}=10^{-9}$ keeping 70 terms to approximate the Hessian (for limited memory) and using at most 20 line search steps in each iteration.
Here \texttt{ftol} and \texttt{gtol} are tolerances used in the stopping criteria of the optimization.
The optimization halts as soon as the absolute or relative change of the objective function between two iterations is smaller than \texttt{ftol}, or the $\ell_1$ norm of the projected gradient is smaller than \texttt{gtol}, or the maximum number of 10,000 iterations has been reached.
We then continue the optimization from the last point with full BFGS with \texttt{gtol}${}=10^{-7}$ for at most 5000 iterations.

A key advancement of our method is that (L)-BFGS does not rely on finite-difference approximations for gradients, but the gradient function can be automatically computed via differential programming.
To this end, we use the \texttt{JAX} package~\cite{jax2018github}, which does not only allow for automatic differentiation, but also includes hardware-specific just-in-time compilation (jit) of Python functions to vastly speed up computations.
We now provide a summary of the overall structure of our implementation in Python syntax.
The full implementation including the found minimizers is publicly available at \url{https://github.com/nikikilbertus/causal-fermion-systems}.

\bigskip

\begin{lstlisting}[language=Python, caption=Structure of the optimization procedure in Python.]
import jax  # popular differentiable programming framework
from scipy.optimize import minimize

# A function to initialize the D real optimization parameters
def initialize_parameters():
    ...
    return parameters

# The causal action as a function of the optimization parameters
def action(parameters):
    ...
    return action

# Automatically get the gradient function for the action
gradient = jax.grad(action)

# Hardware specific just-in-time compilation for speed-up
fast_action = jax.jit(action)
fast_gradient = jax.jit(gradient)

# Initialize the D optimization parameters (randomly)
initial_parameters = initialize_parameters()

# Assume a BFGS implementation is available requiring
# the objective, gradient, and initial guess
minimal_parameters = minimize(
    fun=fast_action,        # objective function
    x0=initial_parameters,  # initial guess
    method="BFGS",          # or "L-BFGS-B"
    jac=fast_gradient,      # gradient function
    options={...}           # parameters for stopping criteria
)
# `minimal_parameters` is a data structure containing the
# minimal parameter vector as the field `x` alongside
# information about convergence

# Compute the action for the minimizer
minimal_action = action(minimal_paramters.x)
\end{lstlisting}

\bibliographystyle{amsplain}

\begin{thebibliography}{10}

\bibitem{cfsweblink}
\emph{Link to web platform on causal fermion systems:
  \href{https://www.causal-fermion-system.com}{www.causal-fermion-system.com}}.

\bibitem{asl2021analysis}
A.~Asl and M.L. Overton, \emph{Analysis of limited-memory bfgs on a class of
  nonsmooth convex functions}, IMA Journal of Numerical Analysis \textbf{41}
  (2021), no.~1, 1--27.

\bibitem{sphere}
L.~B\"auml, F.~Finster, H.~von~der Mosel, and D.~Schiefeneder, \emph{Singular
  support of minimizers of the causal variational principle on the sphere},
  arXiv:1808.09754 [math.CA], Calc. Var. Partial Differential Equations
  \textbf{58} (2019), no.~6, 205.

\bibitem{baydin2018automatic}
A.G. Baydin, B.A. Pearlmutter, A.A. Radul, and J.M. Siskind, \emph{Automatic
  differentiation in machine learning: a survey}, arXiv:1502.05767 [cs.SC],
  Journal of machine learning research \textbf{18} (2018).

\bibitem{bierler}
J.~Bierler, \emph{{Numerische Untersuchung kausaler Variationsprinzipien}},
  Masterarbeit Mathematik, Universit\"at Regensburg (2019).

\bibitem{bjorken}
J.D. Bjorken and S.D. Drell, \emph{Relativistic {Q}uantum {M}echanics},
  McGraw-Hill Book Co., New York, 1964.

\bibitem{jax2018github}
J.~Bradbury, R.~Frostig, P.~Hawkins, M.J. Johnson, C.~Leary, D.~Maclaurin,
  G.~Necula, A.~Paszke, J.~Vander{P}las, S.~Wanderman-{M}ilne, and Q.~Zhang,
  \emph{{JAX}: composable transformations of {P}ython+{N}um{P}y programs},
  http://github.com/google/jax, version 0.2.5 (2018).

\bibitem{conway+sloane}
J.H. Conway and N.J.A. Sloane, \emph{{Sphere Packings, Lattices and Groups}},
  third ed., Grundlehren der Mathematischen Wissenschaften [Fundamental
  Principles of Mathematical Sciences], vol. 290, 1999.

\bibitem{small}
A.~Diethert, F.~Finster, and D.~Schiefeneder, \emph{Fermion systems in discrete
  space-time exemplifying the spontaneous generation of a causal structure},
  arXiv:0710.4420 [math-ph], Int.\ J.\ Mod.\ Phys. A \textbf{23} (2008),
  no.~27/28, 4579--4620.

\bibitem{discrete}
F.~Finster, \emph{A variational principle in discrete space-time: Existence of
  minimizers}, arXiv:math-ph/0503069, Calc. Var. Partial Differential Equations
  \textbf{29} (2007), no.~4, 431--453.

\bibitem{continuum}
\bysame, \emph{Causal variational principles on measure spaces},
  arXiv:0811.2666 [math-ph], J. Reine Angew. Math. \textbf{646} (2010),
  141--194.

\bibitem{cfs}
\bysame, \emph{The {C}ontinuum {L}imit of {C}ausal {F}ermion {S}ystems},
  arXiv:1605.04742 [math-ph], Fundamental Theories of Physics, vol. 186,
  Springer, 2016.

\bibitem{dice2018}
\bysame, \emph{Causal fermion systems: {D}iscrete space-times, causation and
  finite propagation speed}, arXiv:1812.00238 [math-ph], J. Phys.: Conf. Ser.
  \textbf{1275} (2019), 012009.

\bibitem{review}
F.~Finster and M.~Jokel, \emph{Causal fermion systems: An elementary
  introduction to physical ideas and mathematical concepts}, arXiv:1908.08451
  [math-ph], {P}rogress and {V}isions in {Q}uantum {T}heory in {V}iew of
  {G}ravity (F.~Finster, D.~Giulini, J.~Kleiner, and J.~Tolksdorf, eds.),
  Birkh\"auser Verlag, Basel, 2020, pp.~63--92.

\bibitem{gaugefix}
F.~Finster and S.~Kindermann, \emph{A gauge fixing procedure for causal fermion
  systems}, arXiv:1908.08445 [math-ph], J. Math. Phys. \textbf{61} (2020),
  no.~8, 082301.

\bibitem{intro}
F.~Finster, S.~Kindermann, and J.-H. Treude, \emph{{C}ausal {F}ermion
  {S}ystems: {A}n {I}ntroduction to {F}undamental {S}tructures, {M}ethods and
  {A}pplications}, \href{https://arxiv.org/abs/2411.06450}{arXiv:2411.06450
  [math-ph]}, to appear in Cambridge Monographs on Mathematical Physics,
  Cambridge University Press, 2025.

\bibitem{dice2014}
F.~Finster and J.~Kleiner, \emph{Causal fermion systems as a candidate for a
  unified physical theory}, arXiv:1502.03587 [math-ph], J. Phys.: Conf. Ser.
  \textbf{626} (2015), 012020.

\bibitem{jet}
\bysame, \emph{A {H}amiltonian formulation of causal variational principles},
  arXiv:1612.07192 [math-ph], Calc. Var. Partial Differential Equations
  \textbf{56:73} (2017), no.~3, 33.

\bibitem{support}
F.~Finster and D.~Schiefeneder, \emph{On the support of minimizers of causal
  variational principles}, arXiv:1012.1589 [math-ph], Arch. Ration. Mech. Anal.
  \textbf{210} (2013), no.~2, 321--364.

\bibitem{guo2018nonsmooth}
J.~Guo and A.S. Lewis, \emph{Nonsmooth variants of {P}owell's {BFGS}
  convergence theorem}, SIAM Journal on Optimization \textbf{28} (2018), no.~2,
  1301--1311.

\bibitem{harris2020numpy}
C.R. Harris, K.J. Millman, S.J. van~der Walt, R.~Gommers, P.~Virtanen,
  D.~Cournapeau, E.~Wieser, J.~Taylor, S.~Berg, N.J. Smith, R.~Kern, M.~Picus,
  S.~Hoyer, M.H. van Kerkwijk, M.~Brett, A.~Haldane, J.~Fern{\'{a}}ndez del
  R{\'{i}}o, M.~Wiebe, P.~Peterson, P.~G{\'{e}}rard-Marchant, K.~Sheppard,
  T.~Reddy, W.~Weckesser, H.~Abbasi, C.~Gohlke, and T.E. Oliphant, \emph{Array
  programming with {NumPy}}, Nature (2020).

\bibitem{hunter2007matplotlib}
J.D. Hunter, \emph{Matplotlib: A 2{D} graphics environment}, Computing in
  Science \& Engineering (2007).

\bibitem{lawson+michelsohn}
H.B. Lawson, Jr. and M.-L. Michelsohn, \emph{Spin {G}eometry}, Princeton
  Mathematical Series, vol.~38, Princeton University Press, Princeton, NJ,
  1989.

\bibitem{lewis2009nonsmooth}
A.S. Lewis and M.L. Overton, \emph{Nonsmooth optimization via {BFGS}},
  Submitted to SIAM J. Optimization (2009), 1--35.

\bibitem{lewis2013nonsmooth}
\bysame, \emph{Nonsmooth optimization via quasi-{N}ewton methods}, Mathematical
  Programming \textbf{141} (2013), no.~1, 135--163.

\bibitem{nocedal2006numerical}
J.~Nocedal and S.~Wright, \emph{Numerical {O}ptimization}, Springer Science \&
  Business Media, 2006.

\bibitem{peskin+schroeder}
M.E. Peskin and D.V. Schroeder, \emph{An {I}ntroduction to {Q}uantum {F}ield
  {T}heory}, Addison-Wesley Publishing Company Advanced Book Program, Reading,
  MA, 1995.

\bibitem{powell1976some}
M.J.D. Powell, \emph{Some global convergence properties of a variable metric
  algorithm for minimization without exact line searches}, Nonlinear
  programming, SIAM-AMS proceedings, vol.~9, 1976.

\bibitem{reed+simon4}
M.~Reed and B.~Simon, \emph{Methods of {M}odern {M}athematical {P}hysics. {IV},
  {A}nalysis of operators}, Academic Press, New York, 1978.

\bibitem{saff+kuijlaars}
E.B. Saff and A.B.J. Kuijlaars, \emph{Distributing many points on a sphere},
  Math. Intelligencer \textbf{19} (1997), no.~1, 5--11.

\bibitem{sloane}
N.J.A. Sloane, \emph{Tables of sphere packings and spherical codes}, IEEE
  Trans. Inform. Theory \textbf{27} (1981), no.~3, 327--338.

\bibitem{snygg}
J.~Snygg, \emph{Clifford {A}lgebra}, Oxford University Press, New York, 1997, A
  computational tool for physicists.

\bibitem{vanrossum2009python}
G.~van Rossum and F.L. Drake, \emph{{Python 3 Reference Manual}}, CreateSpace,
  2009.

\bibitem{welch}
L.~Welch, \emph{Lower bounds on the maximum cross correlation of signals
  ({Corresp}.)}, IEEE Transactions on Information Theory \textbf{20} (1974),
  no.~3, 397--399, Conference Name: IEEE Transactions on Information Theory.

\end{thebibliography}

\providecommand{\bysame}{\leavevmode\hbox to3em{\hrulefill}\thinspace}
\providecommand{\MR}{\relax\ifhmode\unskip\space\fi MR }
\providecommand{\MRhref}[2]{%
  \href{http://www.ams.org/mathscinet-getitem?mr=#1}{#2}
}
\providecommand{\href}[2]{#2}

\end{document}